# Optical Manipulation of the Charge Density Wave state in RbV$_3$Sb$_5$

*Yuqing Xing[1†], Seokjin Bae[1†], Ethan Ritz[2], Fan Yang[2], Turan Birol[2], Andrea N. Capa Salinas[3], Brenden R. Ortiz[3], Stephen D. Wilson[3], Ziqiang Wang[4], Rafael M. Fernandes[5], Vidya Madhavan[1*]*

[1] Department of Physics and Materials Research Laboratory, University of Illinois Urbana-Champaign, Urbana, IL 61801, USA.

[2] Department of Chemical Engineering and Materials Science, University of Minnesota, Minneapolis, Minnesota 55455, USA

[3] Materials Department, University of California Santa Barbara, Santa Barbara, California 93106, USA

[4] Department of Physics, Boston College, Chestnut Hill, Massachusetts 02467, USA

[5] School of Physics and Astronomy, University of Minnesota, Minneapolis, Minnesota 55455, USA

**Broken time-reversal symmetry in the absence of spin order indicates the presence of unusual phases such as orbital magnetism and loop currents[1–4]. The recently discovered kagome superconductors AV$_3$Sb$_5$ (A = K, Rb, or Cs)[5,6], hosting an exotic charge-density wave (CDW) state, have emerged as strong candidates for this phase. The idea that the CDW breaks time-reversal symmetry[7–14] is however being intensely debated due to conflicting experimental data[15–17]. In this work we use laser-coupled scanning tunneling microscopy (STM) to study RbV$_3$Sb$_5$. STM data shows that the CDW breaks rotational and mirror symmetries. By applying linearly polarized light along high-symmetry directions, we show that the relative intensities of the CDW peaks can be reversibly switched, implying a substantial electro-striction response, indicative of strong non-linear electron-phonon coupling. A similar CDW intensity switching is observed with perpendicular magnetic fields, which implies an unusual piezo-magnetic response that, in turn, requires time-reversal symmetry-breaking. We show that the simplest CDW that satisfies these constraints and reconciles previous seemingly contradictory experimental data is an out-of-phase combination of bond charge order and loop currents that we dub congruent CDW flux phase. Our laser-STM data opens the door to the possibility of dynamic optical control of complex quantum phenomenon in correlated materials.**

†These authors contributed equally to this work
*Correspondence to: vm1@illinois.edu

Kagome lattice compounds provide an ideal platform to study novel states resulting from the interplay between topology, geometric frustration, and electron correlations[18–21]. This class of materials includes magnetic metals such as Fe$_3$Sn$_2$, Co$_3$Sn$_2$S$_2$, and TbMn$_6$Sn$_6$, which host massive Dirac fermions[22–24], magnetic topological Weyl points[25,26], and Chern magnetism[27], as well as the recently discovered non-magnetic superconductors $AV_3Sb_5$ (where A=K, Rb, Cs)[5,6], which exhibit exotic correlated and topological states[28–32]. Besides superconductivity, the $AV_3Sb_5$ compounds host a CDW phase that not only breaks additional lattice symmetries such as rotation[15] and mirror[7], but may also break time-reversal (TR) symmetry[7–14]. The latter is particularly intriguing since broken TR symmetry not driven by spins suggests orbital magnetism with staggered circulating orbital currents[7], resembling states suggested for many decades in the context of quantum anomalous Hall states[1] and correlated systems[2–4]. In the case of $AV_3Sb_5$, TR symmetry-breaking was indicated by scanning tunneling microscopy measurements that detected a magnetic-field sensitive CDW intensity pattern[7]; muon spin-rotation (μSR) experiments that showed an increase in the relaxation rate below the CDW transition temperature[8–10]; and optical measurements that reported the onset of a spontaneous polar Kerr rotation and circular dichroism at the CDW transition[11,12,14]. These observations have motivated various theoretical studies of the so-called flux-CDW phase, which relies on electronic correlations as the driving force behind this exotic instability[33–38]. The scenario of broken TR symmetry has however been challenged by contradictory STM measurements[15] and by Sagnac interferometer measurements that did not find spontaneous Kerr rotation in the CDW state[16,17]. Pinning down the nature of the CDW phase is vitally important not only to determine if the exotic flux-CDW phase exists, but also because superconductivity in these materials emerges deep within the CDW phase.

In this work we study the electric and magnetic field response of the CDW phase of the kagome superconductor $RbV_3Sb_5$ with a laser light coupled STM. $RbV_3Sb_5$ avoids the complexities associated with the co-existence of different types of bond distortions that are seen in the more commonly studied $CsV_3Sb_5$ compound[39–41]. The crystal structure consists of the V-Sb kagome layer sandwiched between two Sb honeycomb layers and separated by an alkaline Rb hexagonal layer (Fig. 1a,b)[5]. Cleavage naturally occurs between the Rb and Sb planes. Since the Rb atoms are weakly bonded to each other and to the Sb plane, the loosely held Rb adatoms can be 'swept away' with the STM tip (Extended Data Fig. 1). The clean Sb surface[42,43] (Fig. 1c and 1d) reveals the Sb honeycomb lattice with a lattice constant of ~ 5.4 Å (Extended Data Fig. 2), consistent with the crystal structure[43–45]. The Fourier transform reveals a rich pattern of charge order and Friedel oscillations (marked as $Q_F$). The three Bragg peaks in the three symmetry-related Γ-M directions are labelled $Q_{B1}$, $Q_{B2}$, and $Q_{B3}$ and the 2a$_0$ x 2a$_0$ CDW peaks along these directions are labelled $Q_1$, $Q_2$, and $Q_3$ respectively. The unidirectional ~ 4a$_0$ CDW[42–44] along the $Q_2$ direction is labeled $Q_{4a_0}$.

A closer look at the FT shown in Fig. 1d reveals a few interesting features. First, the intensity of the $Q_2$ peak (henceforth labelled $I_2$) is significantly higher than the intensity of both $Q_1$ ($I_1$) and $Q_3$ ($I_3$). This can be seen by comparing the line profiles along three Bragg peak directions as shown in Fig. 1e, and indicates the breaking of threefold rotational symmetry of the 2a$_0$ x 2a$_0$ CDW, consistent with previous literature[7,15,43,44,46]. Theoretically, this rotation symmetry breaking has been attributed to either a staggered stacking of the CDW along the c-axis or to an admixture of the bond-CDW with a flux phase[35,36,38,47]. Second, we find that $I_1$ and $I_3$ are also different, which indicates that vertical and diagonal mirror symmetries are also broken such that all three CDW peaks have different intensities. We will show later in the paper that the difference between $I_1$ and $I_3$ is likely due to small residual strain. The lack of mirror planes perpendicular to the kagome layer gives rise to two different senses of handedness in the CDW peak intensities, which were referred to as "chiral" in previous STM studies[7,44], when they appeared to be switchable by applying a c-axis magnetic field. To understand the origin of this remarkable phenomenon, we study the response of the CDW to a different type of electromagnetic probe, i.e., light.

A schematic of our laser-STM setup is shown in Fig. 2a. More details about laser parameters, optics setup, and laser-STM experimental procedures can be found in Methods and Extended Data Fig. 3. Figure 2 shows the FT of the topography and the 2a$_0$ x 2a$_0$ CDW peak intensities before and after illumination. In this study, the laser electric field ($E$) is oriented along the in-plane Γ-M directions pointing towards either $Q_1$ or $Q_3$ (we henceforth label light with $E \parallel Q_1$ as $E_1$ and $E \parallel Q_3$ as $E_3$). We note here that the directions $E_1$ and $E_3$ are along the honeycomb Sb-Sb nearest neighbor direction in Q-space and therefore along the V-V nearest neighbor directions in real space. Before illumination, $I_1 > I_3$ as shown in Fig. 2c. This intensity difference is emphasized in the 3D plot of just the CDW peaks (Fig. 2d). Strikingly, we find that after illumination with $E_3$ (Fig. 2e,f), $I_1 < I_3$. This light-induced CDW intensity change is reversible. Illuminating the sample with $E_1$ reverses the sequence of intensities again, leading to $I_1 > I_3$

(Fig. 2 g,h). This finding is independent of bias voltage and can be seen in the dI/dV maps as shown in Extended Data Fig. 4 and Fig. 5. Due to the inherent broken rotational symmetry of the CDW, $I_2$ always remains the strongest so for the rest of this paper we will not discuss $I_2$.

To demonstrate the robustness and repeatability of the light induced switching, we show the result of a series of laser illuminations with light polarized either along $E_1$ or $E_3$ in Figure 3a. Here we plot the relative intensity $I_r = (I_1 - I_3)/((I_1 + I_3)/2)$ where a positive/negative $I_r$ (marked in red or blue) corresponds to a stronger $I_1$ or $I_3$ respectively (see Extended Data Fig. 6 for details on the method used to quantify the intensities). Plotting $I_r$ rather than the absolute intensity values allows us to eliminate arbitrary intensity changes in the overall FT between measurements. Initially, before laser illumination, $I_1 > I_3$, leading to a positive $I_r$. After laser illumination with $E_3$, $I_r$ becomes negative meaning that now $I_1 < I_3$. Amazingly, as we track $I_r$ through a sequence of light pulses, we find a one-to-one correspondence between the sign of $I_r$ and the direction of the electric field, as shown in Fig. 3a (also see Fig. 3b for the full statistics of 33 different illuminations in an arbitrary sequence). This means that shining light along $E_1$ or $E_3$ always makes the intensity of the corresponding CDW peak stronger. There are three other important points to make. First, the observed laser control unequivocally proves that the intensity differences are a feature of the sample and not measurement artifacts. Second, the switching is independent of the number of pulses in an illumination and occurs even within the single pulse limit as shown in Fig. 3f. Finally, the laser induced switching requires a critical fluence of ~ 0.2 mJ/cm² (Fig. 3e).

CDW order typically goes hand-in-hand with lattice distortions. Remarkably, we are able to directly observe such intertwining by tracking the positions of the Bragg and CDW peaks as the direction of the electric field is switched. To see this, we extract the ratio between the magnitude of the Bragg vectors i.e., $Q_r = |Q_{B1}|/|Q_{B3}|$ as an indicator of the change in the relative lattice constants after laser illumination (see Extended Data Fig. 8 for further details on how $Q_B$ is determined). Fig. 3c shows that $Q_r$ displays the same pattern as the relative CDW intensity $I_r$ shown in Fig. 3a. In particular, the ratio $Q_r$ increases when illuminating with $E_1$ and decreases when illuminating with $E_3$. This means that when $I_1$ increases, $Q_{B1}$ increases and when $I_3$ increases, $Q_{B3}$ increases. This trend is preserved with different peak identification methods, as shown in Extended Data Fig. 10.

Concomitantly with the change in the Bragg momentum, the ratio between the CDW peak positions also changes (Extended Data Fig. 11) such that the 2a₀ x 2a₀ CDW remains commensurate after laser illumination (Fig. 3d). In other words, while the absolute value of the CDW wave-vector changes, its coordinates with respect to the reciprocal lattice vectors remain unchanged. These combined observations suggest a strong electron-phonon coupling in this system[48] and establish that the light-induced intensity-change can be attributed to a sizable electro-striction response of the CDW state.

The surprising laser-induced switching of CDW intensities observed here is similar to the previously reported magnetic field switching of the relative CDW intensities[7,44]. To obtain a comprehensive picture, it is important to ascertain if a similar magnetic field induced switching occurs in our samples. Shown in Fig. 4a-b are the FT of the topography under a magnetic field of -2T. First, even with this new sample and tip and in a different STM, we observe the same anisotropic amplitudes of the CDW peaks along the three Γ-M directions. Second, comparing $I_1$ and $I_3$ under -2T (Fig. 4a-b, 4e-f) and +2T fields (Fig. 4c-d), we see that the relative CDW intensities switch in a reversible manner. The robustness of this magnetic-field-induced switching, is illustrated by the CDW response to a sequence of out-of-plane magnetic fields (Fig. 4g). Importantly, the sign-change in $I_r$ is accompanied by a change in the relative length of the Bragg vector $\boldsymbol{Q}_r$ (Fig. 4h) as in the case of light induced intensity changes. Our data thus not only provide independent confirmation of the CDW intensity reversal in magnetic fields seen previously[7,44], but also uncover that the reversal is accompanied by field-induced anisotropic strain.

The ability to manipulate the relative intensities of the CDW peaks with linearly polarized light and magnetic field gives us valuable information on the CDW order parameter such as its relation to time-reversal symmetry. To extract this information, we employ a phenomenological analysis of the experimental results, which relies only on symmetry constraints and not on specific microscopic mechanisms. Specifically, we consider a general flux CDW phase described by a three-component CDW order parameter $\boldsymbol{L} = (L_1, L_2, L_3)$ with wave-vectors $\boldsymbol{Q}_{L_1} = \left(\frac{1}{2}, 0, \frac{1}{2}\right)$, $\boldsymbol{Q}_{L_2} = \left(0, \frac{1}{2}, \frac{1}{2}\right)$, and $\boldsymbol{Q}_{L_3} = \left(-\frac{1}{2}, +\frac{1}{2}, \frac{1}{2}\right)$, corresponding to distortions of the V bonds that increase the size of the unit cell by $2 \times 2 \times 2$. We also include in the description of the flux CDW phase a time-reversal symmetry-breaking (TRSB) CDW order parameter $\boldsymbol{\Phi} = (\Phi_1, \Phi_2, \Phi_3)$ with in-plane wave-vectors $\boldsymbol{Q}_{M_1} = \left(\frac{1}{2}, 0, 0\right)$, $\boldsymbol{Q}_{M_2} = \left(0, \frac{1}{2}, 0\right)$, and $\boldsymbol{Q}_{M_3} = \left(-\frac{1}{2}, +\frac{1}{2}, 0\right)$, corresponding to V orbital magnetism ("loop currents"). Our STM data gives direct access to the intensities $I_1, I_2$ and $I_3$ of the CDW peaks at $\boldsymbol{Q}_1, \boldsymbol{Q}_2$, and $\boldsymbol{Q}_3$ respectively.

In the absence of electromagnetic fields, X-ray data on RbV$_3$Sb$_5$ reveal an orthorhombic crystal structure, which is symmetric with respect to a vertical mirror[40]. As a result, two of the CDW peaks are expected to have the same magnitude – more concretely, $I_1 = I_3 \neq I_2$ if the mirror plane is chosen to include the $\boldsymbol{Q}_2$ wave-vector. In terms of the CDW order parameters, this implies $|L_1| = |L_3| \neq |L_2|$. The fact that in our STM data $I_1$ and $I_3$ are slightly different even in the absence of electromagnetic fields is likely due to small residual strain. As our laser measurements have shown, the CDW in these kagome systems is highly sensitive to lattice strain. Our goal is to elucidate what further constraints are imposed to these CDW order parameters by the response of the CDW intensities on the electromagnetic fields.

Let us consider the light-induced response first. Incident light with polarization $\boldsymbol{E}_1$ or $\boldsymbol{E}_3$ results in a lattice distortion accompanied by $I_1 \neq I_3$ (Fig. 5a) such that the mirror plane along $\boldsymbol{Q}_2$ is broken. This observation suggests that the system has a non-zero electro-striction tensor element $\chi^{\text{es}}_{xyxy}$. Such a tensor element is allowed for any CDW configuration with $|L_1| = |L_3| \neq |L_2|$, and as such does not impose additional constraints on the type of CDW order, including on the possible emergence of TRSB. Note that, while symmetry allows a non-zero $\chi^{\text{es}}_{xyxy}$, it says nothing about its magnitude. The sizable effect of the 1.2 eV optical pulse on the CDW peaks, which are associated with low-energy zone-corner phonon modes, is indicative of a strong electron-phonon non-linear coupling[48].

The response of the CDW peaks to the out-of-plane magnetic field $\boldsymbol{B}_z$ imposes even more severe constraints on the CDW order parameters $\boldsymbol{L}$ and $\boldsymbol{\Phi}$. The fact that the relative peak intensities $I_r$ and the ratio of Bragg vectors $\boldsymbol{Q}_r$ switch with the direction of $\boldsymbol{B}_z$ implies a non-zero piezo-magnetic tensor element $\chi^{\text{pm}}_{xyz}$ – or, in other words, that an out-of-plane magnetic field induces an in-plane shear lattice distortion (Fig. 5b). First, this type of piezo-magnetic response necessarily implies that the CDW must break time-reversal symmetry, i.e., that not only $|L_1| = |L_3| \neq |L_2|$, but also $|\Phi_1| = |\Phi_3| \neq |\Phi_2|$. Moreover, as we show in the group theoretical analysis in Methods, it implies a non-zero $L_2$ and a vanishing $\Phi_2$, so that the system does not have a macroscopic magnetic dipole, and, remarkably, a non-trivial relative phase of $\pi$ between the non-zero components of the two order parameters, $\text{sign}(L_1 L_3 \Phi_1 \Phi_3) = -1$. This relative phase has a simple interpretation in terms of the symmetries of the bond-order pattern, described by $L$, and the loop-current pattern, described by $\Phi$. To illustrate this, consider the simpler case of a single kagome layer. Separately, the triple-Q bond-order pattern and the double-Q loop-current pattern each have an in-plane twofold rotation axis (Extended Data Fig. 14a-c). The key point is that the rotation axes do not match if the aforementioned relative phase is trivial (Fig. 5c), but match when it is non-trivial (Fig. 5d). It is only in the latter case that the combined flux CDW phase possesses an in-plane twofold rotation axis. This is essential for the piezo-magnetic response observed experimentally in the absence of macroscopic magnetization. Upon application of an out-of-plane magnetic field, the two-fold rotation axis is broken, resulting in a shear strain. We dub this unusual phase where the rotation axes of both the bond-order pattern and the loop-current pattern match a congruent flux CDW phase.

A schematic of the response of the CDW, and the corresponding lattice distortions, to electric and magnetic fields is summarized in Fig. 5a,b and 5e. In the absence of the fields and of residual strain, the peaks at the two CDW wave-vectors are equivalent ($|L_1| = |L_3|$), resulting in the schematic double-well free-energy shown by the red curve in Fig. 5e. The application of light with polarization along a particular direction ($E_1$ for instance, Fig. 5a) changes the free energy landscape in Fig. 5e and favors one of the two CDW order parameters. This results in an increase in $I_1$ and a corresponding decrease in the lattice parameter in this direction. This finite electro-striction response is likely mediated by light-activated phonons and non-linear electron-phonon coupling. Similarly, a magnetic field towards the sample surface (Fig. 5b) decreases the lattice parameter along the direction of the Bragg vector $Q_{B3}$ resulting in an increase in the CDW intensity at $Q_3$. Taken together, via its non-trivial and significant electro-striction and piezo-magnetic responses, the system can switch from one local minimum to the other within the free energy landscape shown in Fig. 5e.

The response of the CDW to electromagnetic fields provides strong constraints on the symmetry of the order parameter. The simplest CDW configuration that satisfies these conditions is the congruent flux phase $L = (|L_1|, L_2, |L_1|)$ and $\Phi = (|\Phi_1|, 0, -|\Phi_1|)$, which has the *magnetic* space group $Cmmm$ (#65.481) and which breaks time reversal symmetry (see group theoretical

analysis in Methods). Note that this CDW configuration can also address several other experimental observations. First, it is consistent with one of the non-magnetic space groups ($Cmmm$, #65) that were found to possibly refine the crystal structure of RbV$_3$Sb$_5$ via X-ray data[40]. Second, it is not only consistent with TRSB seen by μSR experiments, but it can also reconcile the seemingly contradictory spontaneous Kerr effect results reported in the literature. While a pristine sample would not display a spontaneous Kerr effect for this CDW configuration, as it does not have a macroscopic magnetic dipole moment, the non-zero piezomagnetic tensor element $\chi_{xyz}^{\text{pm}}$ implies that unintentional shear strain in the sample would generate an out-of-plane magnetic moment, which in turn would give rise to a spontaneous Kerr effect. Lastly, our laser-STM study paves the way to an in-situ optical manipulation of strain and symmetry breaking in quantum materials.

**Acknowledgements**

The laser-STM studies were supported by the Gordon and Betty Moore Foundation's EPiQS initiative through Grant No. GBMF9465. This material is based upon work supported by the U.S. Department of Energy Office of Science National Quantum Information Science Research Centers as part of the Q-NEXT center, which supported the work of S. B. and V. M and provided partial support for laser-STM development. Funding for sample growth was provided via the UC Santa Barbara NSF Quantum Foundry funded via the Q-AMASE-i program under award DMR-1906325. A.C.S. acknowledges support from the Eddlemam Center for Quantum Innovation at UC Santa Barbara. E.R, F.Y. and T.B. were supported by the NSF CAREER grant DMR-2046020. R.M.F. was supported by the Air Force Office of Scientific Research under Award No. FA9550-21-1-0423. Z. W. is supported by the U.S. Department of Energy, Basic Energy Sciences (Grant No. DE-FG02-99ER45747) and by Research Corporation for Science Advancement (Cottrell SEED Award No. 27856).

**Author Contributions:**

Y. X., S. B., and V. M. conceived the project. Y.X. and S.B. conducted laser-STM measurement. S. B. constructed the laser-STM setup and designed the laser experiment. Y.X. conducted STM study under magnetic field. A.C.S, B.R.O. and S.D.W. provided $RbV_3Sb_5$ samples used in this study. E.R, F.W., T. B., Z.W., and R. M. F. conducted group theory analysis and theoretical interpretation of the data. Y.X., S.B., and V. M. performed data analysis and wrote the manuscript with input from all the authors.

**Competing interests:** The authors declare that they have no competing interests.

**Data Availability:** All of the data for the main figures will be uploaded to the Illinois Databank.

**Figure 1**

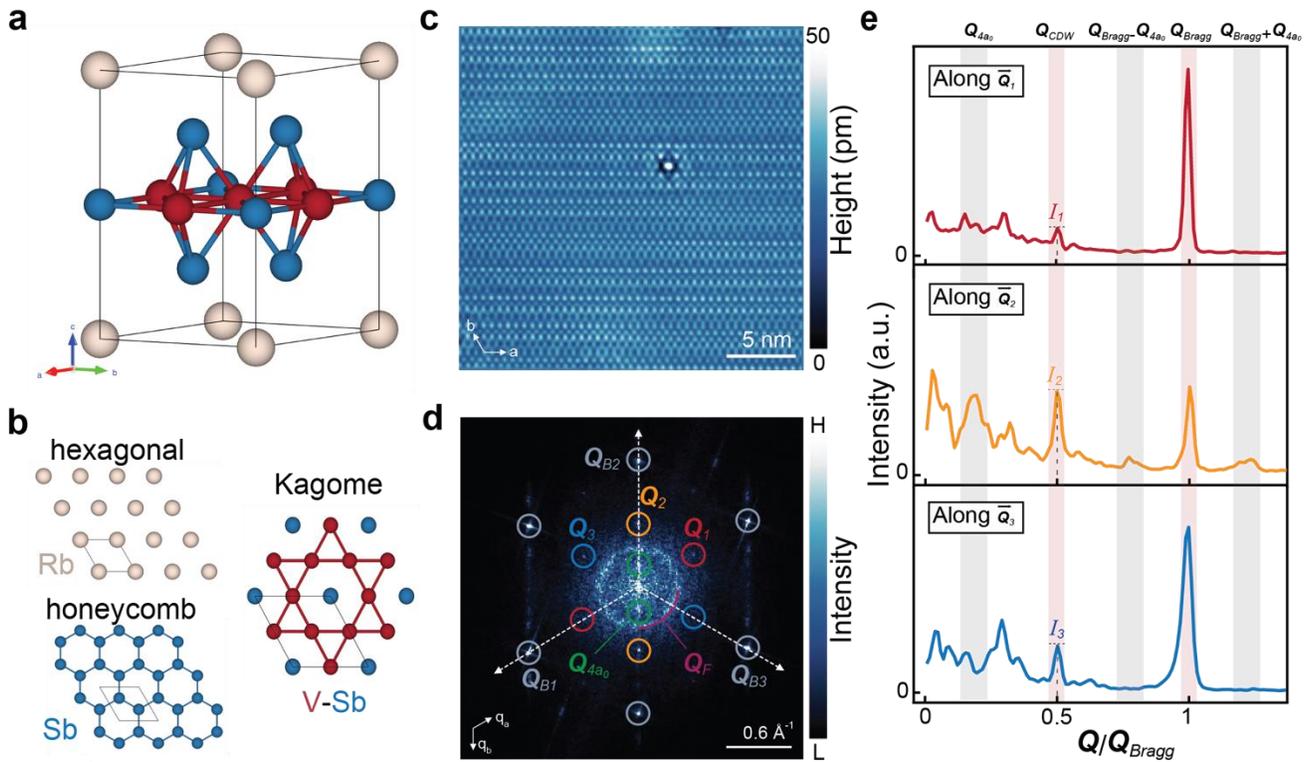

**Fig. 1 | Sb surface identification of RbV$_3$Sb$_5$ and CDW peak intensities.**

**a**, Unit cell of RbV$_3$Sb$_5$. **b**, Top view of different terminations, showing the hexagonal Rb layer, the honeycomb Sb layer and the kagome V-Sb layer. **c**, Topographic image (25 nm x 25 nm) of the Sb layer ($V_{sample}$ = -100 mV, $I_{tunneling}$ =175 pA). **d**, FT image of the Sb surface, showing the wave vectors $Q_{1-3}$ associated with the 2$a_0$ x 2$a_0$ CDW, $Q_{4a_0}$ associated with unidirectional charge order and $Q_F$ associated with isotropic defect scattering related interference patterns. The three Bragg peaks along the Γ-M directions are labelled $Q_{B1-B3}$. **e**, Comparison of linecuts in (d) along the $Q_1$ (top), $Q_2$ (middle) and $Q_3$ (bottom) directions. $Q_{4a_0}$ and its satellite Bragg peaks are only prominent along the $Q_2$ direction, while the 2$a_0$ x 2$a_0$ $Q_{1-3}$ CDW peaks (labelled on top as $Q_{CDW}$) show different intensities in the three directions.

**Figure 2**

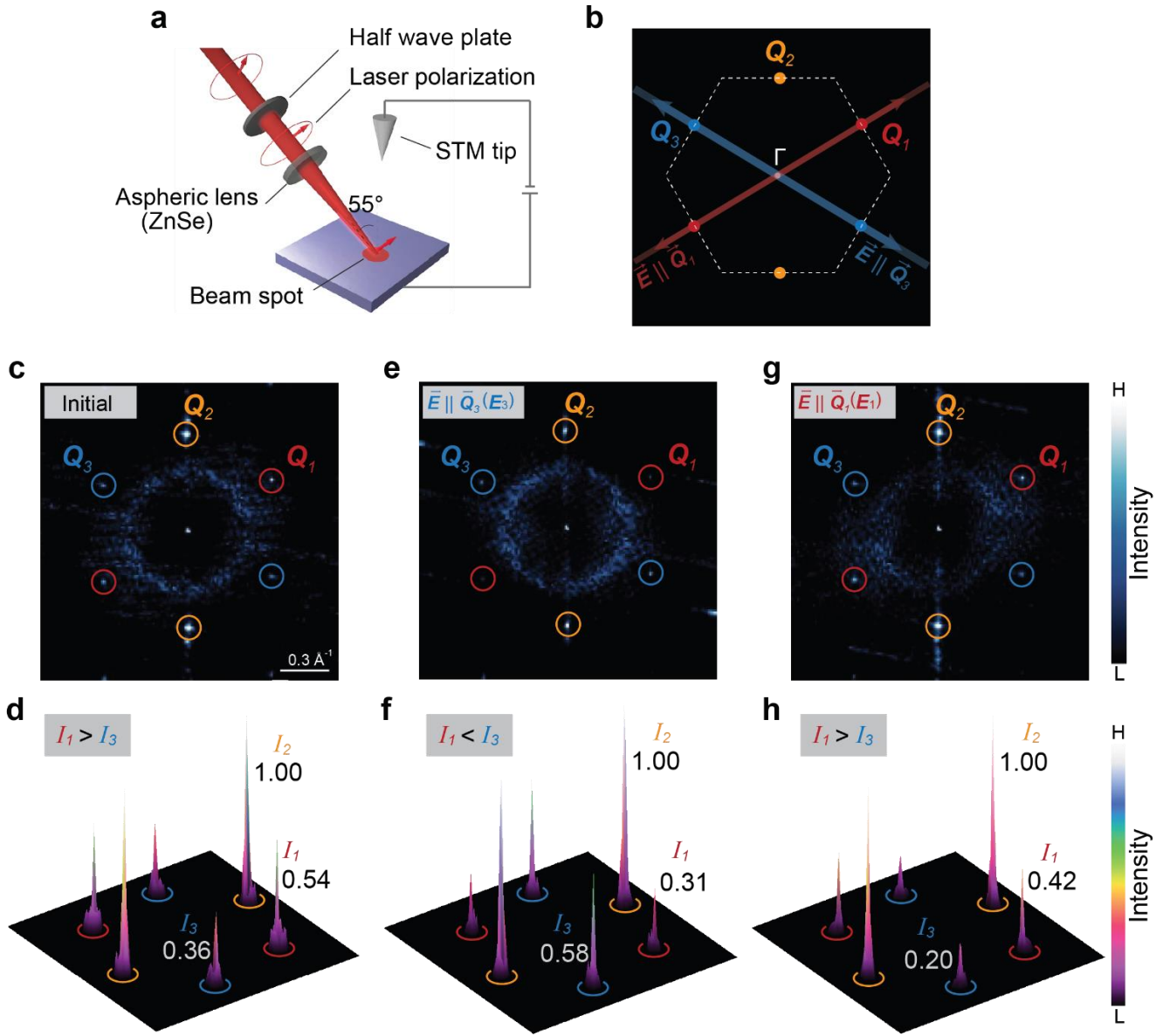

**Fig. 2 | Light-induced switching of the intensity order in 2a₀ x 2a₀ CDW**.

**a**, Schematic illustration of light illumination and subsequent STM measurement on RbV$_3$Sb$_5$. A burst (2 x 10$^5$ shots at 100 kHz repetition rate) of linearly polarized, ultrafast (250 fs), near-infrared (1025 nm) laser pulses at a fluence of 0.39 mJ/cm$^2$ (unless otherwise mentioned) is used to illuminate the sample surface. **b**, Schematic FT image focusing on the 2a₀ x 2a₀ CDW intensity distribution. The red and blue double arrows denote the polarization direction of the laser beam. **c**, FT of topographic image before light illumination ($V_s$ = -100 mV, $I_t$ = 175 pA). **d**, 3D view of the $2a_0 \times 2a_0$ CDW intensity peaks of (c), showing $I_1 > I_3$. To isolate the CDW peaks the circular quasiparticle interference pattern at the center is masked. **e**, FT of topographic image of the same region after light illumination with linear polarization parallel to the $E_3$ direction showing that the intensities have changed upon light illumination such that $I_3 > I_1$. **f**, 3D view of (e) **g**, FT of topographic image of the same region after light illumination with the linear polarization parallel to $E_1$. **h**, 3D view of (g), demonstrating that the intensity order has now switched back to $I_1 > I_3$.

**Figure 3**

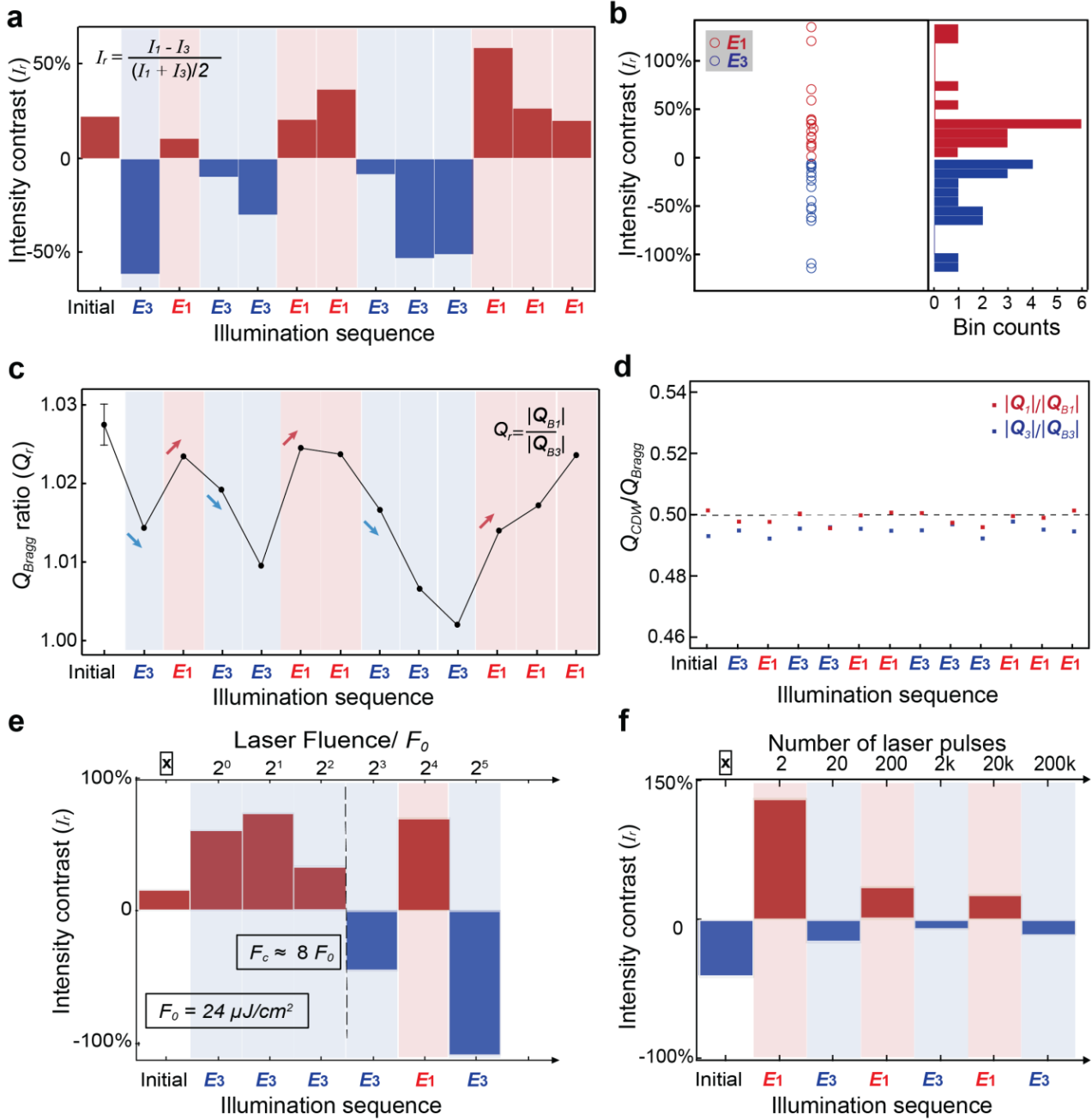

**Fig. 3 | Characterization of the light-induced switching of the relative CDW intensity.**

**a**, Arbitrary illumination sequence with laser polarization along either $E_1$ or $E_3$. The color of the solid bar shows the magnitude and sign of the relative intensity ($I_r$) as defined in the inset. The color of the background indicates the direction of illumination. **b**, Statistics of $I_r$ for $E_1$ illuminations (red circles) and $E_3$ illuminations (blue circles). 33 bin counts in total are displayed in the left sub-panel. The bin counts for a given intensity contrast range are displayed in the right sub-panel. **c**, Plot of the Bragg peak ratio ($Q_r$) for the arbitrary illumination sequence shown in (a). The error bar represents the full range variation of $Q_r$ in five sequential measurements of $Q_r$ with circularly polarized pulsed light, which we have determined does not influence the sign of the CDW intensity contrast. **d**, $Q_{CDW}/Q_{Bragg}$ ratio along $Q_1$ (red dots) and $Q_3$ directions (blue dots), showing that the $2a_0 \times 2a_0$ CDW remains commensurate during the illumination sequence. **e**, Laser pulse fluence dependence of the switching behavior, showing that the threshold fluence to trigger the switching is ~ 0.2 mJ/cm². **f**, Dependence of the switching on the number of shots, showing that the CDW intensity order can be switched even with two pulses.

**Figure 4**

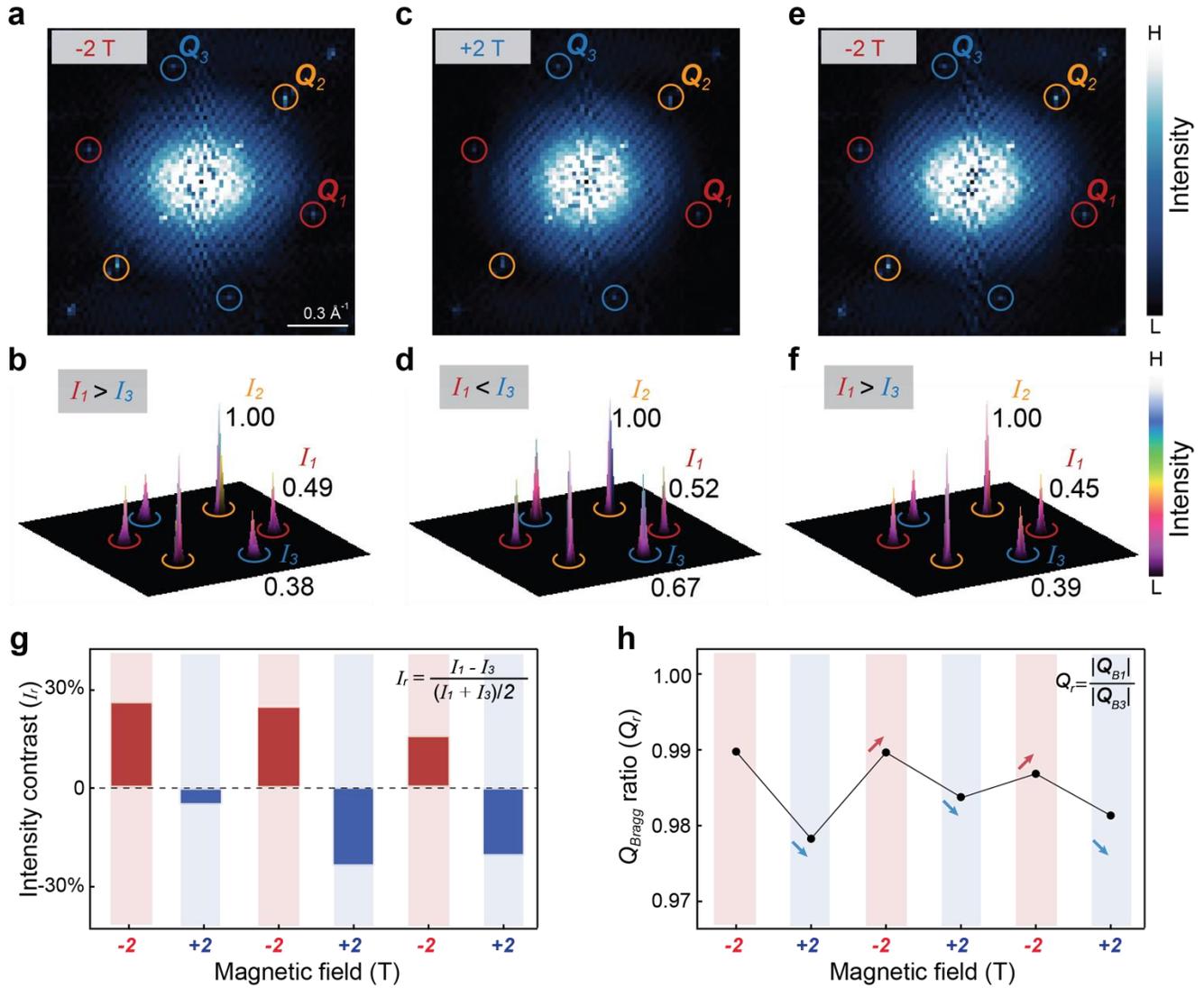

**Fig. 4 | Magnetic-field-induced switching of the relative CDW intensity.**

**a,c,e**, FT of topography image of Sb surface at -2 T/+2 T/-2 T showing a switching of the relative intensities of the CDW peaks depending on the field direction. **b,d,f**, 3D view of the $2a_0 \times 2a_0$ CDW peaks of (a,c,e). The intensity of the $Q_2$ CDW peak is normalized to 1.00. **g**, A plot of $I_r$ as the magnetic field direction is switched in a sequential fashion from negative to positive. The solid colors show the magnitude and sign of the relative intensity ($I_r$) as defined in the inset. The background color indicates the direction of the magnetic field. **h**, Bragg peak ratios ($Q_r$) for the magnetic field sequence shown in (g).

**Figure 5**

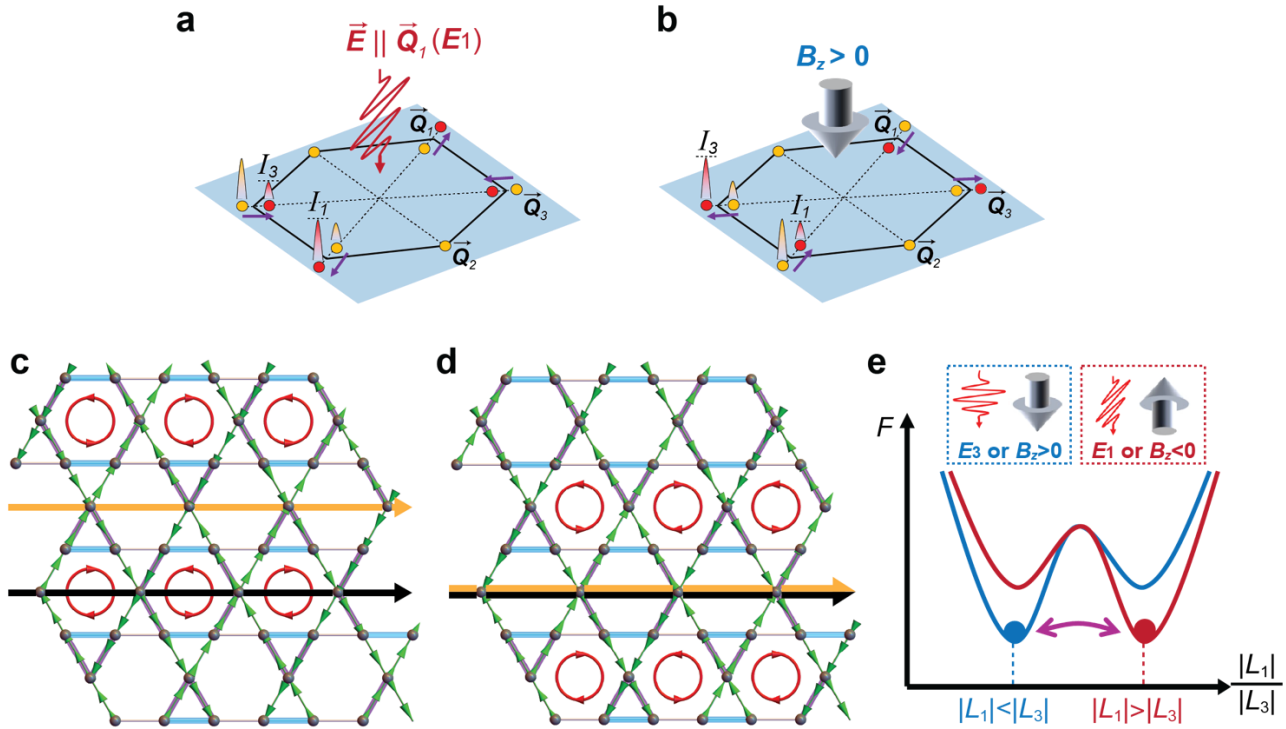

**Fig. 5 | The congruent flux-CDW phase.**

**a**, Schematic illustration of the electro-striction response of the system. The wavy line denotes the optical pulse. The wave amplitude direction denotes its polarization direction (here, the $Q_1$ direction). The position of the red dots and purple arrows represents the response of the Bragg momentum upon illumination. **b**, Schematic illustration of the piezo-magnetic response. The grey 3D arrow denotes the direction of the external magnetic field (here, positive field). **c,d**, An overlayed view of the spatial pattern of Vanadium bond order $L$ (sky-blue and purple colored bond) and loop current order $\Phi$ (green arrows). The black/orange lines represent the in-plane 2-fold rotation axes of the bond order/loop current, respectively. The red circle represents the closed current loops. (**c**) For the case when the non-zero components of $\Phi$ and $L$ have a trivial relative phase (0), $L$ and $\Phi$ do not share the same 2-fold rotation axis. (**d**) For the case when the relative phase is non-trivial ($\pi$), $L$ and $\Phi$ share the same rotation axis, resulting in a congruent flux-CDW phase. **e**, Schematic free-energy landscape of the system in terms of the components of the CDW order parameter corresponding to wave-vectors $Q_1$ and $Q_3$ under external electromagnetic fields. Note that the curves are schematic and do not necessarily have the same shape for the cases of an applied electric and magnetic field.

# Methods and Extended Data Figures for

Optical Manipulation of the Charge Density Wave state in RbV$_3$Sb$_5$

## Table of Contents

**Methods**

### Single crystal growth of RbV$_3$Sb$_5$

Single crystals of RbV$_3$Sb$_5$ were synthesized from Rb (ingot, Alfa 99.75%), V (powder, Sigma 99.9%, purified in an HCl and Ethanol mixture), and Sb (shot, Alfa 99.999%) using a modified self-flux method. Inside an argon glovebox with oxygen and moisture levels < 0.5 ppm, elemental reagents were weighed out to composition Rb$_{20}$V$_{15}$Sb$_{120}$, and loaded into a tungsten carbide vial to be milled for 60 min in a SPEX 8000D mill. Resulting powder was loaded into 2 mL alumina (Coorstek) crucibles and sealed inside steel tubes under argon. Sample was heated to 1000 °C at 200 °C/hour, soaked at that temperature for 12 hours, then cooled down to 900 °C at 5 °C, to finally be slow-cooled to 500 °C at a rate of 1°C /hour. Single crystals of RbV$_3$Sb$_5$ were obtained and mechanically extracted, with typical dimensions of 2 mm x 2 mm x 200 um.

### Scanning tunneling microscopy/spectroscopy

The RbV$_3$Sb$_5$ samples used in the experiments cleaved *in situ* at ~90 K and immediately transferred to an STM chamber. All STM experiments were performed in an ultrahigh vacuum (1×10$^{-10}$ mbar). Laser-coupled STM measurements were conducted under 4 K and Perpendicular magnetic field coupled STM measurements were studied under 1.7 K. All the scanning parameter (setpoint voltage and current) of the STM topographic images are listed in the captions of the figures. Unless otherwise noted, the differential conductance (dI/dV) spectra were acquired by a standard lock-in amplifier at a modulation frequency of 913.1 Hz. STM tip made from tungsten was fabricated via electrochemical etching.

### Laser parameters, optics setup, and experimental procedure for laser-STM

The optical excitation used in this experiment is given by 2 x 10$^5$ burst shots of laser pulse train at 100 kHz repetition rate and at 0.39 mJ/cm$^2$ fluence (otherwise mentioned), 1025 nm wavelength (1.21 eV), and 250 fs pulse width. The laser beam is generated by Yb:KGW RA laser system (PHAROS PH2-10W, Light Conversion). The pulse train timing is controlled by an arbitrary function generator (ARB Rider 2182, Active Technologies) which directly triggers a pulse picker inside the RA laser system. As seen from the detailed layout of the laser-STM setup in Extended Data Fig. 3a. The beam passes through a half-wave plate for the polarization direction control. The half-wave plate rotates the polarization to a direction of interest. Then the beam is guided to the tip-sample junction area inside the USM1200LL STM (Unisoku) through a CaF2 viewport (MPF) and ZnSe aspheric lens (Avantier Inc.) which acts as an objective lens. The focal length of the ZnSe lens at 1025 nm is 12.43 mm. The alignment of the beam spot and the STM tip-sample junction is monitored by an optical microscope setting which consists of another ZnSe lens, N-BK7 plano-convex lens, and a CMOS camera. During the alignment check, the laser beam is switched to continuous pulse train mode and pass through an OD > 2.0 ND filter to prevent melting of the sample surface. The alignment is adjusted by moving the lens piezo stages where ZnSe lenses are mounted. The incident angle of the beam to the surface normal direction of the sample is 55 degrees. The beam is focused to an elliptic spot (due to the incidence angle) with a size estimated to be 49 $\mu$m x 68 $\mu$m (1/e$^2$ diameter).

The procedure used for the laser-STM experiment is illustrated in Extended Data Fig. 3b. When sample is illuminated by the laser burst shots, the tip is retracted along the surface normal by ≈ 70 $\mu$m in order to prevent changes to the tip during illumination. After the burst illumination, the tip is reapproached. The lateral shift after this procedure is less than 10 nm, so it is easy to locate the cleaned Sb surface area and study the same area after every burst.

**Group theoretical analysis on the response of CDW order parameter under electric and magnetic field**

Our starting point to interpret the STM data is a "real" CDW order parameter $\mathbf{L} = (L_1, L_2, L_3)$, corresponding to in-plane and out-of-plane distortions of the V-V bonds, and an "imaginary" CDW order parameter $\mathbf{\Phi} = (\Phi_1, \Phi_2, \Phi_3)$, which corresponds to in-plane orbital magnetic currents involving the V states. In terms of the *P6/mmm* space group (#191) of AV$_3$Sb$_5$, $\mathbf{L}$ transforms as the $L_2^-$ irreducible representation (irrep). Thus, each component $L_i$ is associated with each wave-vector $\mathbf{Q}_{L_1} = \left(\frac{1}{2}, 0, \frac{1}{2}\right)$, $\mathbf{Q}_{L_2} = \left(0, \frac{1}{2}, \frac{1}{2}\right)$, and $\mathbf{Q}_{L_3} = \left(-\frac{1}{2}, +\frac{1}{2}, \frac{1}{2}\right)$, respectively; note that the three different $\mathbf{Q}_{L_i}$ are related by a threefold rotation. On the other hand, $\mathbf{\Phi}$ transforms as the $mM_2^+$ irrep (where $m$ indicates that it is odd under time reversal) and thus each component $\Phi_i$ has wave-vector $\mathbf{Q}_{M_1} = \left(\frac{1}{2}, 0, 0\right)$, $\mathbf{Q}_{M_2} = \left(0, \frac{1}{2}, 0\right)$, and $\mathbf{Q}_{M_3} = \left(-\frac{1}{2}, +\frac{1}{2}, 0\right)$. Note that $mM_2^+$ is the simplest type of loop-current order that can be constructed out of the low-energy van Hove singularities (vHs), corresponding to intra-orbital (or intra-vHs) order[1]. Because in DFT both the phonon mode $L_2^-$ and the phonon mode $M_1^+$ irrep are unstable, we also consider the real CDW order parameter $\boldsymbol{M} = (M_1, M_2, M_3)$ for completeness[2]. Note that $\boldsymbol{M}$ transforms as the $M_1^+$ irrep and, like $\boldsymbol{\Phi}$, also has wave-vector $\boldsymbol{Q}_{M_1} = \left(\frac{1}{2}, 0, 0\right)$, $\boldsymbol{Q}_{M_2} = \left(0, \frac{1}{2}, 0\right)$, and $\boldsymbol{Q}_{M_3} = \left(-\frac{1}{2}, +\frac{1}{2}, 0\right)$. In the notation adopted here, the reciprocal lattice vectors are given by:

$$\boldsymbol{G}_1 = \frac{2\pi}{a}\left(1, \frac{1}{\sqrt{3}}, 0\right)$$

$$\boldsymbol{G}_2 = \frac{2\pi}{a}\left(0, \frac{2}{\sqrt{3}}, 0\right)$$

$$\boldsymbol{G}_3 = \frac{2\pi}{c}(0,0,1) \tag{1}$$

We start by writing down the full Landau free-energy for the coupled order parameters, $F = F_L + F_\Phi + F_M + F_{L\Phi} + F_{LM} + F_{\Phi M}$. We have, for the free terms (see also Ref. [1]):

$$F_L = \frac{a_L}{2}\sum_i L_i^2 + \frac{u_L}{4}\left(\sum_i L_i^2\right)^2 + \frac{\lambda_L}{4}(L_1^2 L_2^2 + L_1^2 L_3^2 + L_2^2 L_3^2)$$

$$F_\Phi = \frac{a_\Phi}{2}\sum_i \Phi_i^2 + \frac{u_\Phi}{4}\left(\sum_i \Phi_i^2\right)^2 + \frac{\lambda_\Phi}{4}(\Phi_1^2 \Phi_2^2 + \Phi_1^2 \Phi_3^2 + \Phi_2^2 \Phi_3^2)$$

$$F_M = \frac{a_M}{2}\sum_i M_i^2 + \frac{\gamma_M}{3} M_1 M_2 M_3 + \frac{u_M}{4}\left(\sum_i M_i^2\right)^2 + \frac{\lambda_M}{4}(M_1^2 M_2^2 + M_1^2 M_3^2 + M_2^2 M_3^2) \tag{2}$$

For the coupled terms, we obtain:

$$F_{L\Phi} = \frac{\kappa_{L\Phi}}{4}(L_1 L_2 \Phi_1 \Phi_2 + L_1 L_3 \Phi_1 \Phi_3 + L_2 L_3 \Phi_2 \Phi_3)$$
$$+ \frac{\lambda_{L\Phi}}{4}(L_1^2 \Phi_1^2 + L_2^2 \Phi_2^2 + L_3^2 \Phi_3^2) + \frac{u_{L\Phi}}{4}\left(\sum_i L_i^2\right)\left(\sum_i \Phi_i^2\right) \tag{3}$$

as well as

$$F_{LM} = \frac{\gamma_{LM}}{3}(L_1 L_2 M_3 + L_1 L_3 M_2 + L_2 L_3 M_1) + \frac{\kappa_{LM}}{4}(M_1 M_2 L_1 L_2 + M_1 M_3 L_1 L_3 + M_2 M_3 L_2 L_3)$$

$$+ \frac{\lambda_{LM}}{4}(M_1^2 L_1^2 + M_2^2 L_2^2 + M_3^2 L_3^2) + \frac{u_{LM}}{4}\left(\sum_i M_i^2\right)\left(\sum_i L_i^2\right) \quad (4)$$

and

$$F_{\Phi M} = \frac{\gamma_{\Phi M}}{3}(\Phi_1 \Phi_2 M_3 + \Phi_1 \Phi_3 M_2 + \Phi_2 \Phi_3 M_1) + \frac{\kappa_{\Phi M}}{4}(M_1 M_2 \Phi_1 \Phi_2 + M_1 M_3 \Phi_1 \Phi_3 + M_2 M_3 \Phi_2 \Phi_3)$$

$$+ \frac{\lambda_{\Phi M}}{4}(M_1^2 \Phi_1^2 + M_2^2 \Phi_2^2 + M_3^2 \Phi_3^2) + \frac{u_{\Phi M}}{4}\left(\sum_i M_i^2\right)\left(\sum_i \Phi_i^2\right) \quad (5)$$

Rather than minimizing the free energy, we look for order parameter configurations that can explain the STM data. For concreteness, we will consider the instabilities driven by the **L** and **Φ** channels, such that the accompanying **M** order parameter follows trivially from the trilinear couplings $\gamma_{LM}, \gamma_{\Phi M}$ above.

The key quantity measured by STM is the relative intensity between the three CDW peaks at wave-vectors $\mathbf{Q}_{M_1} = (1/2, 0, 0)$, $\mathbf{Q}_{M_2} = (0, 1/2, 0)$, and $\mathbf{Q}_{M_3} = (-1/2, +1/2, 0)$ which correspond to $\boldsymbol{Q}_1, \boldsymbol{Q}_2$, and $\boldsymbol{Q}_3$ in the main text. It can be conveniently described in terms of the following two-component "vector" (see, e.g. Ref. [2]):

$$\psi_M = \begin{pmatrix} M_1^2 + M_3^2 - 2M_2^2 \\ \sqrt{3}(M_3^2 - M_1^2) \end{pmatrix} \quad (6)$$

which transforms as the $\Gamma_5^+$ irrep. Indeed, if all three peaks have the same intensity, $M_1^2 = M_2^2 = M_3^2$, $\psi_M \sim (0 \ 0)^T$. On the other hand, if one of the peaks is stronger or weaker than the other two, say $M_2^2 \neq M_1^2 = M_3^2$, then $\psi_M \sim (1 \ 0)^T$. The other two possibilities, corresponding to $M_1^2 \neq M_2^2 = M_3^2$ and $M_3^2 \neq M_1^2 = M_2^2$, give $\psi_M \sim (1 \ \sqrt{3})^T$ and $\psi_M \sim (1 \ -\sqrt{3})^T$. Any other non-zero value of $\psi_M$ corresponds to three different peak intensities. Importantly, the quantity $\psi_M$ has the same transformation properties as:

$$\psi_L = \begin{pmatrix} L_1^2 + L_3^2 - 2L_2^2 \\ \sqrt{3}(L_3^2 - L_1^2) \end{pmatrix} \tag{7}$$

Therefore, hereafter, we will focus only on $\psi_L$. The three values associated with the case where one peak is stronger or weaker than the other two equivalent peaks correspond to:

$$\psi_L^1 \sim \begin{pmatrix} 1 \\ 0 \end{pmatrix}, \quad \psi_L^2 \sim \begin{pmatrix} 1 \\ \sqrt{3} \end{pmatrix}, \quad \psi_L^3 \sim \begin{pmatrix} 1 \\ -\sqrt{3} \end{pmatrix} \tag{8}$$

Clearly, $\psi_L^1$, $\psi_L^2$, and $\psi_L^3$, are related by a 120° rotation. Therefore, if one peak becomes different than the other two, three-fold rotational symmetry is broken. Note, however, that there remains a vertical mirror plane – or, equivalently, an in-plane two-fold rotation axis $C_2'$. These results can also be obtained by noticing that the in-plane strain components $\varepsilon_{xx} - \varepsilon_{yy}$ and $\varepsilon_{xy}$ also transform as the same $\Gamma_5^+$ irrep as $\psi_M$ and $\psi_L$:

$$\varepsilon_\parallel = \begin{pmatrix} \varepsilon_{xx} - \varepsilon_{yy} \\ -2\varepsilon_{xy} \end{pmatrix} \tag{9}$$

where $\varepsilon_{ij} \equiv (\partial_i u_j + \partial_j u_i)/2$ and $u$ is the displacement vector. The coupling to the **L** order parameter is given by:

$$F_{L\varepsilon} = \alpha_1 \left[ (\varepsilon_{xx} - \varepsilon_{yy})(L_1^2 + L_3^2 - 2L_2^2) - 2\sqrt{3}\varepsilon_{xy}(L_3^2 - L_1^2) \right] \tag{10}$$

where $\alpha_1$ is some coupling constant. Hence, if $\psi_L$ is in one of the three configurations described by Eq. (8), the system is in an orthorhombic phase, whereas for any other non-zero $\psi_L$ values, the system is in a monoclinic phase.

We start by analyzing the STM results in the absence of electromagnetic fields. Previous x-ray measurements have shown that the system is in an orthorhombic phase[3]. For concreteness, we choose the configuration corresponding to $\psi_L^1$, i.e. $|L_1| = |L_3| \neq L_2$. From Eq. (10), we see that while $\varepsilon_{xx} \neq \varepsilon_{yy}$, there is no shear strain, $\varepsilon_{xy} = 0$. Moreover, in this case, the vertical mirror plane contains the reciprocal lattice vectors $\boldsymbol{G_2}$ and $\boldsymbol{G_3}$ and is thus perpendicular to the $\hat{q}_x$ axis in momentum space, $\hat{q}_x \equiv (1, -1/2, 0)$. We note that, in the experiments, the CDW peaks corresponding to $L_1$ and $L_3$ are slightly different. We attribute this behavior to residual shear strain $\varepsilon_{xy}$ present in the sample which, via the coupling in Eq. (10), induces $L_3^2 \neq L_1^2$.

Consider now the application of an in-plane electric field $E_\parallel = (E_x, E_y)$. Since it transforms as the $\Gamma_6^-$ irrep, its coupling to the $L$ order parameter has a similar form as Eq. (10):

$$F_{LE} = \alpha_2 \left[ (E_x^2 - E_y^2)(L_1^2 + L_3^2 - 2L_2^2) - 2\sqrt{3} E_x E_y (L_3^2 - L_1^2) \right] \tag{11}$$

Parametrizing $E_\parallel = E_0 (\cos\theta, \sin\theta)$, we have:

$$F_{LE} = \alpha_2 E_0^2 \left[ \cos 2\theta (L_1^2 + L_3^2 - 2L_2^2) - \sqrt{3} \sin 2\theta (L_3^2 - L_1^2) \right] \tag{12}$$

In the experiments, the electric field is applied either along the $Q_{M_1} \equiv (1/2, 0, 0) = \frac{\pi}{a}\left(1, \frac{1}{\sqrt{3}}, 0\right)$ direction or the $Q_{M_3} \equiv (-1/2, +1/2, 0) = \frac{\pi}{a}\left(-1, \frac{1}{\sqrt{3}}, 0\right)$ direction. In terms of their polar angles $\theta_1$ and $\theta_3$, these two directions are related according to $\theta_3 = \pi - \theta_1$. As a result, $\cos 2\theta_1 = \cos 2\theta_3$ but $\sin 2\theta_1 = -\sin 2\theta_3$. Consequently, according to Eq. (12), application of an electric field along these two directions leads to a splitting of the two CDW peaks, $L_3^2 - L_1^2 = \pm \delta L^2$, with opposite signs for a field along $\theta_1$ and a field along $\theta_3$. In terms of the strain tensor, the coupling in Eq. (12) can be written as:

$$F_{\varepsilon E} = \alpha_2' \left[ (E_x^2 - E_y^2)(\varepsilon_{xx} - \varepsilon_{yy}) + 4 E_x E_y \varepsilon_{xy} \right] \tag{13}$$

From the definition of the electro-striction response tensor, $\varepsilon_{ij} = \gamma_{ijkl} E_j E_k$, we readily identify $\alpha_2' \propto \gamma_{66}$ (in Voigt notation) or, using the main text notation for $\gamma_{ijkl}$, $\alpha_2' \propto \chi_{xyxy}^{\text{es}}$. This tensor element is allowed (i.e. not enforced to be zero by symmetry) for both orthorhombic groups *Fmmm* and *Cmmm*, as well as for the hexagonal group *P6/mmm*. Therefore, the STM observation of a switch in the relative intensity of the CDW peaks at $Q_{M_1}$ and $Q_{M_3}$ when the electric field direction is switched between the two directions does not require that the CDW breaks time-reversal symmetry.

Consider now the application of an out-of-plane magnetic field $B_z$, which transforms as the $m\Gamma_2^+$ irrep. The STM data shows that the CDW peaks $L_1^2$ and $L_3^2$ become different, and that the relative intensity flips sign when $B_z$ changes sign. According to Eq. (10), a difference in $L_1^2$ and $L_3^2$ triggers a shear distortion $\varepsilon_{xy} \neq 0$, whose sign depends on the sign of the magnetic field. Now, the piezomagnetic response tensor is defined as $\varepsilon_{ij} = \Lambda_{ijk} B_k$. Hence, this observation implies a non-zero tensor element $\Lambda_{63} \neq 0$ (in mixed Voigt notation) or, using the notation of the main text for $\Lambda_{ijk}$, a non-zero $\chi^{\text{pm}}_{xyz}$.

Clearly, $\Lambda_{63} \neq 0$ requires that time-reversal symmetry is broken, which suggests that the CDW phase must contain a non-zero orbital magnetic density-wave order parameter $\boldsymbol{\Phi}$. To identify under which configurations of $\mathbf{L}$ and $\boldsymbol{\Phi}$ a non-zero $\Lambda_{63}$ is allowed, we write down their coupling to $B_z$ and $\varepsilon_\parallel$ (to leading order):

$$F_{L\Phi, B\varepsilon} = \alpha_3 B_z \left[ \sqrt{3}\, \varepsilon_{xy} L_2 (\Phi_1 L_3 - \Phi_3 L_1) - \frac{1}{2}(\varepsilon_{xx} - \varepsilon_{yy})(\Phi_1 L_2 L_3 - 2\Phi_2 L_1 L_3 + \Phi_3 L_1 L_2) \right] \quad (14)$$

We can readily identify:

$$\Lambda_{63} = \alpha_3 \sqrt{3}\, L_2 (\Phi_1 L_3 - \Phi_3 L_1) \quad (15)$$

Recall that the constraints on $\mathbf{L}$ so far are $|L_1| = |L_3| \neq L_2$. A non-zero $\Lambda_{63}$ further requires $|\Phi_1| = |\Phi_3|$ and $\text{sign}(L_1 L_3 \Phi_1 \Phi_3) = -1$, i.e. it imposes constraints not only on the magnitude of the loop-current order parameter, but also on the relative phase between the non-zero components of $\mathbf{L}$ and $\boldsymbol{\Phi}$. The second term in Eq. (14) also imposes the constraint $\Phi_2 = 0$, otherwise the CDW phase in the orthorhombic phase would spontaneously induce a non-zero magnetic dipole moment (since $\varepsilon_{xx} \neq \varepsilon_{yy}$), which is not observed experimentally.

The result of this analysis is that the CDW order parameter configuration consistent with the STM experimental observations is the one with $\mathbf{L} = (L_0, L_0', L_0)$ and $\boldsymbol{\Phi} = (\Phi_0, 0, -\Phi_0)$. From the leading-order couplings to the order parameter $\mathbf{M}$ in $F_{LM}$ and $F_{\Phi M}$, we also find the subsidiary order $\mathbf{M} = (M_0, M_0', M_0)$, with $M_0 \sim L_0 L_0'$ and $M_0' \sim L_0^2$. We checked that this configuration results in the magnetic space group *Cmmm* (#65.481), which indeed allows for a non-zero $\Lambda_{63}$. We extended this analysis in a systematic way to other configurations of $\mathbf{L}$ and $\boldsymbol{\Phi}$, as shown in the Extended Data Table 1, as well as to cases in which $\boldsymbol{\Phi}$ transforms as other irreps, but did not find any configurations that lead to an orthorhombic magnetic space group that has $\Lambda_{63} \neq 0$.

The large number of Landau coefficients in the free energy of Eqs. (2)-(5) makes it difficult to fully sample the ground states realized in parameter space. Nevertheless, we can identify which terms play a major role in favoring the $\mathbf{L} = (L_0, L'_0, L_0)$ and $\mathbf{\Phi} = (\Phi_0, 0, -\Phi_0)$ order via a qualitative analysis. Specifically, while $\lambda_{L\Phi} < 0$ and $u_{L\Phi} < 0$ favor coexistence between $\mathbf{L}$ and $\mathbf{\Phi}$ orders, the sign of $\kappa_{L\Phi}$ selects between the $(L_0, L'_0, L_0)$ and $(\Phi_0, 0, -\Phi_0)$ combination ($\kappa_{L\Phi} > 0$), which displays the desired piezomagnetic properties, and the $(L_0, L'_0, L_0)$ and $(\Phi_0, 0, \Phi_0)$ combination ($\kappa_{L\Phi} < 0$), which does not display the desired piezomagnetic properties. We also would like to emphasize that the $(L_0, L'_0, L_0)$ and $(\Phi_0, 0, -\Phi_0)$ order is one of the unique ground states emerging from arbitrary combinations of $\mathbf{L}$ and $\mathbf{\Phi}$, as we verified via the ISOTROPY software. Finally, it is worth noting that, in Ref. 3, where a simplified free energy for only the real CDW order parameters was studied, a rich phase diagram with various non-trivial combinations of the components of $\mathbf{L}$ and $\mathbf{M}$ was obtained.

**Extended References**

1. Christensen, M. H., Birol, T., Andersen, B. M. & Fernandes, R. M. Loop currents in AV$_3$Sb$_5$ kagome metals: Multipolar and toroidal magnetic orders. *Phys. Rev. B* **106**, 144504 (2022).
2. Christensen, M. H., Birol, T., Andersen, B. M. & Fernandes, R. M. Theory of the charge-density wave in AV$_3$Sb$_5$ kagome metals. *Phys. Rev. B* **104**, 214513 (2021).
3. Kautzsch, L. *et al.* Structural evolution of the kagome superconductors AV$_3$Sb$_5$ (A = K, Rb, and Cs) through charge density wave order. *Phys. Rev. Mater.* **7**, 024806 (2023).

**Extended Data Table 1**

| $L_2^-$ bond-order CDW | $mM_2^+$ loop-current CDW | Magnetic Space Group | Macroscopic Magnetization | Nonzero Components of Piezomagnetic Tensor |
|---|---|---|---|---|
| $L_1 = L_2 \neq L_3$ | $\Phi_1 = -\Phi_2, \Phi_3 = 0$ | Cmmm (#65.481) | none | $yzx = zyx$<br>$xzy = zxy$<br>$xyz = yxz$ |
| $L_1 = -L_2, L_3 = 0$ | $\Phi_1 = \Phi_2, \Phi_3 = 0$ | $C_A mmm$ (#65.490) | none | none |
| $L_1 = -L_2, L_3 = 0$ | $\Phi_1 = -\Phi_2, \Phi_3 = 0$ | $C_A ccm$ (#66.500) | none | none |
| $L_1 = L_2 \neq L_3$ | $\Phi_1 = \Phi_2 \neq \Phi_3$ ($\Phi_3$ may be zero) | Cm'm'm (#65.485) | $M_z$ | $xxz$<br>$yyz$<br>$zzz$<br>$yzy = zyy$<br>$xzx = zxx$ |
| $L_1 \neq L_2 \neq L_3$ | $\Phi_1 \neq \Phi_2 \neq \Phi_3$ | P2/m (#10.42) | $M_z$ | $yyz$<br>$zzz$<br>$xxz$<br>$zxy$<br>$zxx$<br>$yxz$<br>$yzy$<br>$yzx$ |

**Extended Data Table 1: Magnetic space groups for different combinations of the bond-order CDW and the loop-current CDW order parameters.**

Of the possibilities considered, only the congruent CDW phase exhibits the required piezo-magnetism in the absence of a macroscopic magnetization. Note that the combination of order parameter $L_1 = L_2 \neq L_3$ and $\Phi_1 = -\Phi_2 \neq \Phi_3$ with $\Phi_3 \neq 0$ reduces the symmetry down to monoclinic P2/m, where none of the order parameters components remain equal.

**Extended Data Fig. 1**

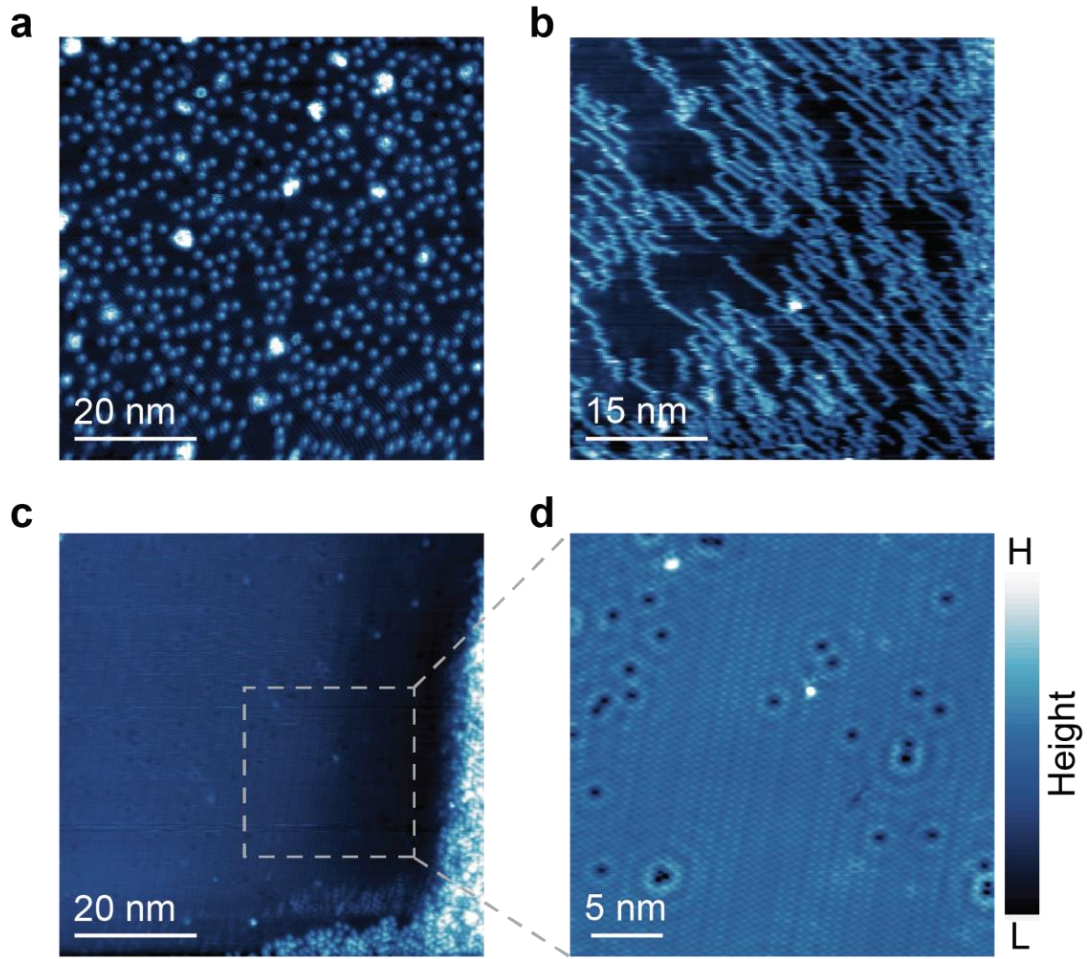

**Extended Data Fig. 1 | Revealing the Sb surface by the sweeping method.**

**a**, Initial topography image (70 nm x 70 nm) showing Sb surface with randomly distributed Rb adatoms on top ($V_s$ = -200 mV; $I_t$ = 100 pA). **b**, 'Sweeping' away surface Rb adatom. Moving adatoms are seen as lines in the topography ($V_s$ = -50 mV; $I_t$ = 1.5 nA). **c**, After adatom manipulation, most of the adatoms move to the corner of the scan frame (70 nm x 70 nm), exposing a clean Sb surface ($V_s$ = -100 mV; $I_t$ = 100 pA). **d**, Zoom-in topography image (30 nm x 30 nm) of dashed square region in (c), showing the honeycomb structure of the Sb surface ($V_s$ = -100 mV; $I_t$ = 500 pA).

**Extended Data Fig. 2**

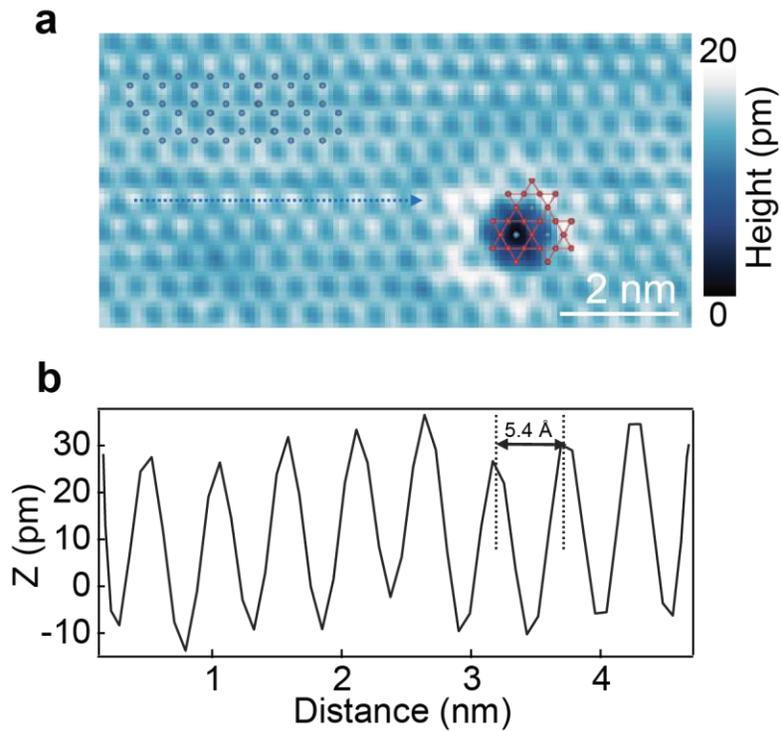

**Extended Data Fig. 2 | High-resolution topography and lattice constant of Sb layer.**

**a**, Atomic-resolution topography of the Sb lattice ($V_s$ = -200 mV; $I_t$ = 100 pA) with an overlap of the crystal structure highlighting the honeycomb Sb lattice (blue) and underlying kagome V lattice (red). **b**, Line profile of dashed blue line in (a), denoting a ~5.4 Å surface atomic spacing corresponding to the Sb lattice.

**Extended Data Fig. 3**

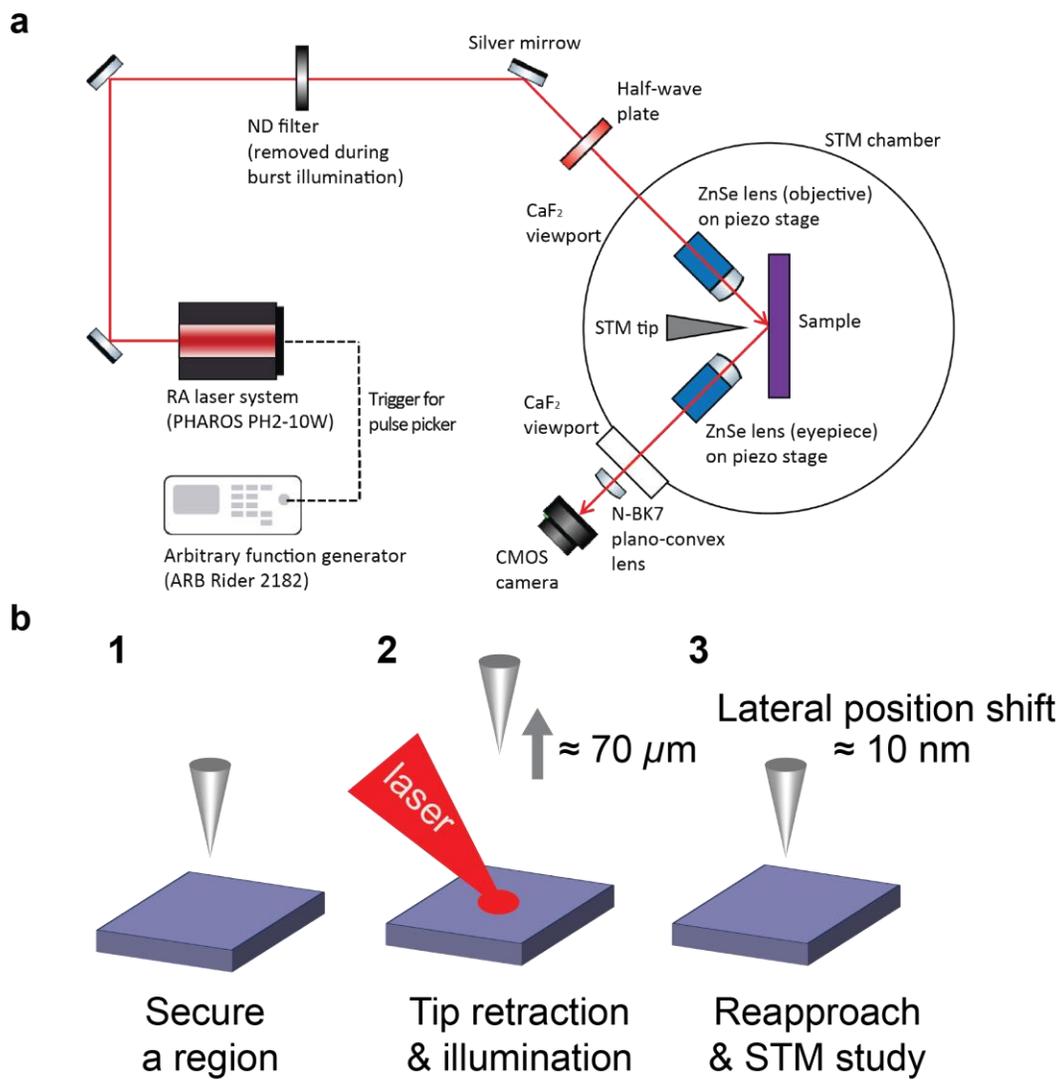

**Extended Data Fig. 3 | a,** Optics layout of the laser STM. **b,** Experimental procedure to measure light-induced changes in CDW intensity.

**Extended Data Fig. 4**

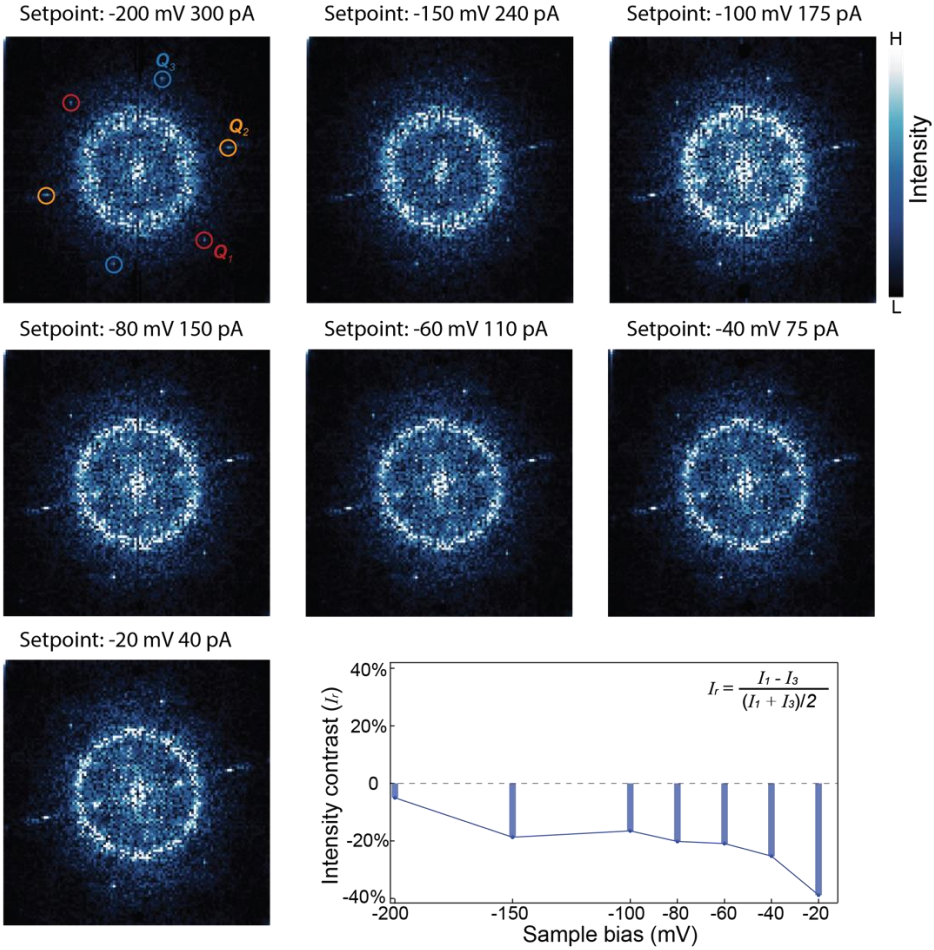

**Extended Data Fig. 4 | Evolution of the CDW intensity ratio at different sample biases.**
For different biases, the sign of $I_r$ remains the same.

**Extended Data Fig. 5**

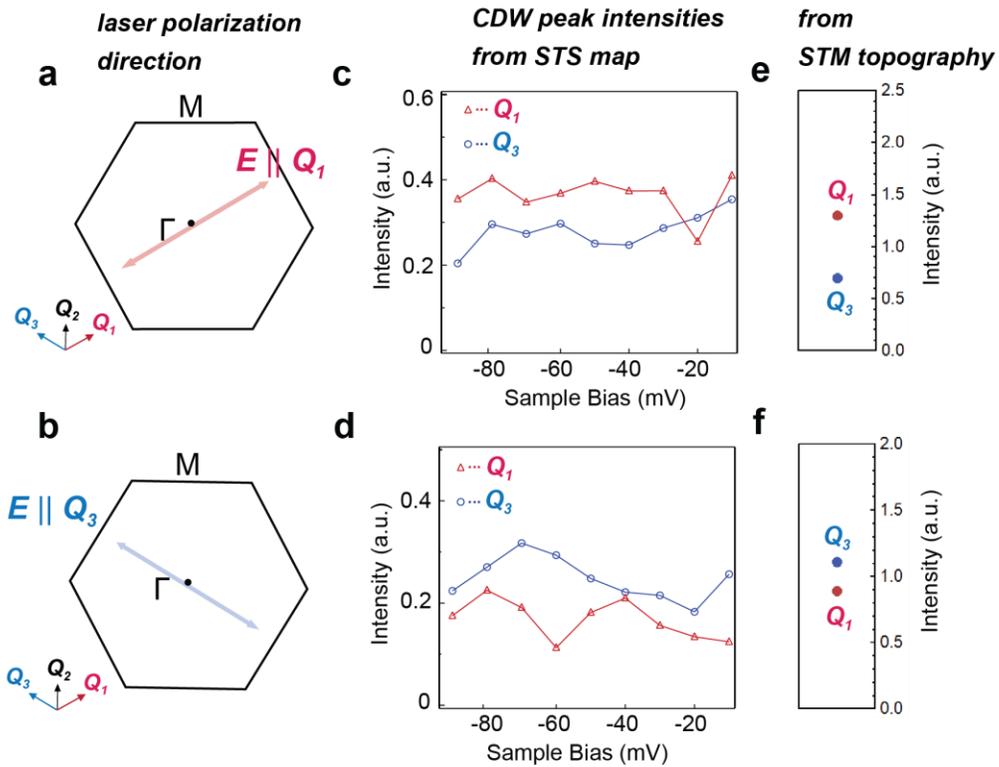

**Extended Data Fig. 5 | Scanning tunneling spectroscopy (STS) maps and STM topography showing the behavior of the CDW intensity upon illumination of linearly polarized light.**

**a-b**, The two directions of laser polarization with respect to the schematic reciprocal lattice. The red and blue double arrows denote the polarization direction of the laser beam. Upon laser illumination along $Q_1$ and $Q_3$ directions, the same switching behaviour of the CDW intensities appears both in the STS map (**c-d**, $V_{bias}$ = -200 mV, $I_{set}$ = 100 pA) and STM topography (**e-f**).

**Extended Data Fig. 6**

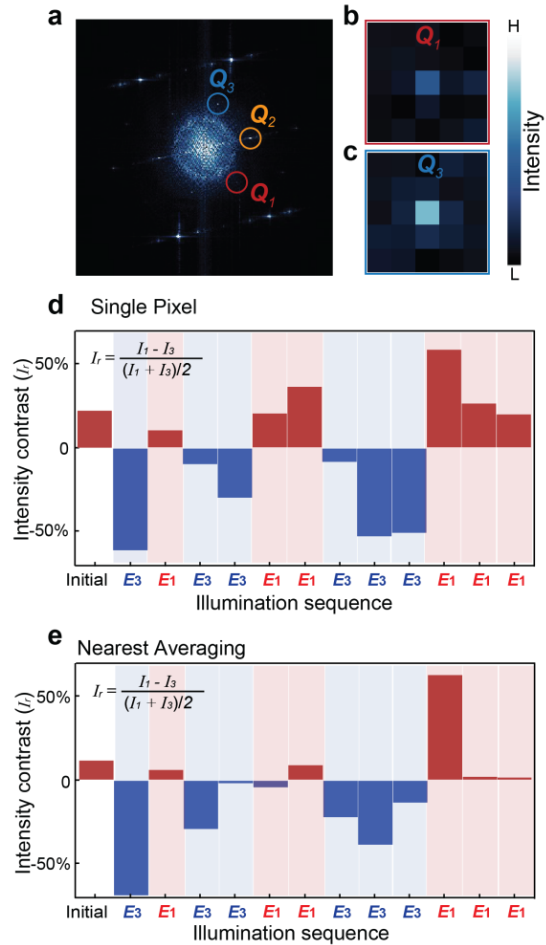

**Extended Data Fig. 6 | Agreement between different methods of quantifying the CDW intensity.**

**a**, Typical FT of the Sb surface. The 2 x 2 CDW peaks along the three directions are labelled $Q_1$ to $Q_3$. **b-c**, Zoomed in FT images showing that the intensities of $Q_1$ and $Q_3$ are mostly localized to a single pixel. **d**, Light-induced switching of the CDW intensity with an arbitrary illumination sequence with laser polarization along either $\boldsymbol{E}_1$ or $\boldsymbol{E}_3$ (same as Fig. 3a), with the CDW intensities determined by the single pixel method. **e**, CDW intensity contrast along the same arbitrary illumination sequence with the nearest averaging methods (5 pixels in total). The trends in the intensity contrast remain the same while the absolute value of the intensity contrast is suppressed in most cases suggesting that using the single pixel method provides better signal.

**Extended Data Fig. 7**

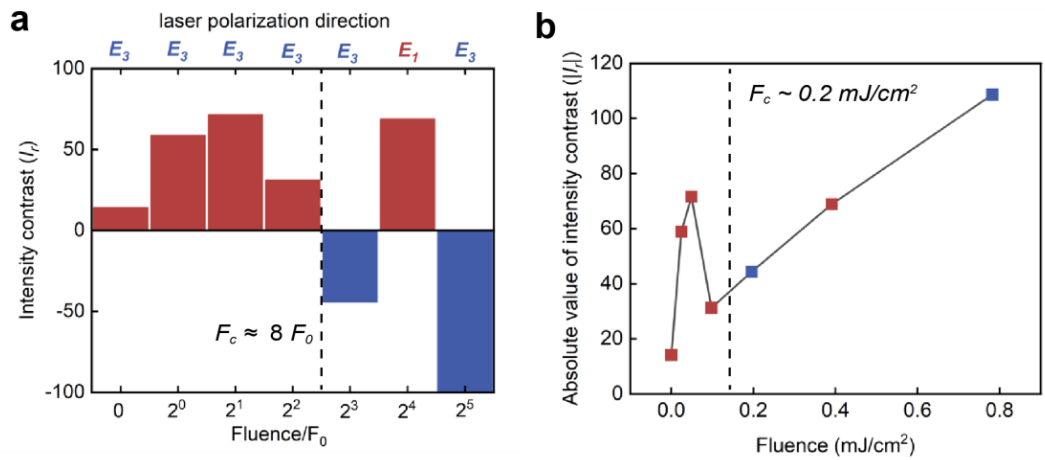

**Extended Data Fig. 7 | Laser fluence dependence of the CDW intensity contrast.**

**a**, Fig. 3e in main text: $I_r = (I_1 - I_3)/2(I_1 + I_3)$ at each illumination with doubling the fluence. Beyond the critical fluence $F_c \approx 8\, F_0$, the sign of $I_r$ starts to change depending on the direction of illumination. **b**, Absolute value of the $I_r$ ($|I_r|$) with respect to the fluence on a linear scale. For fluence > $F_c$, the fluence and $|I_r|$ start to show a proportional relation.

**Extended Data Fig. 8**

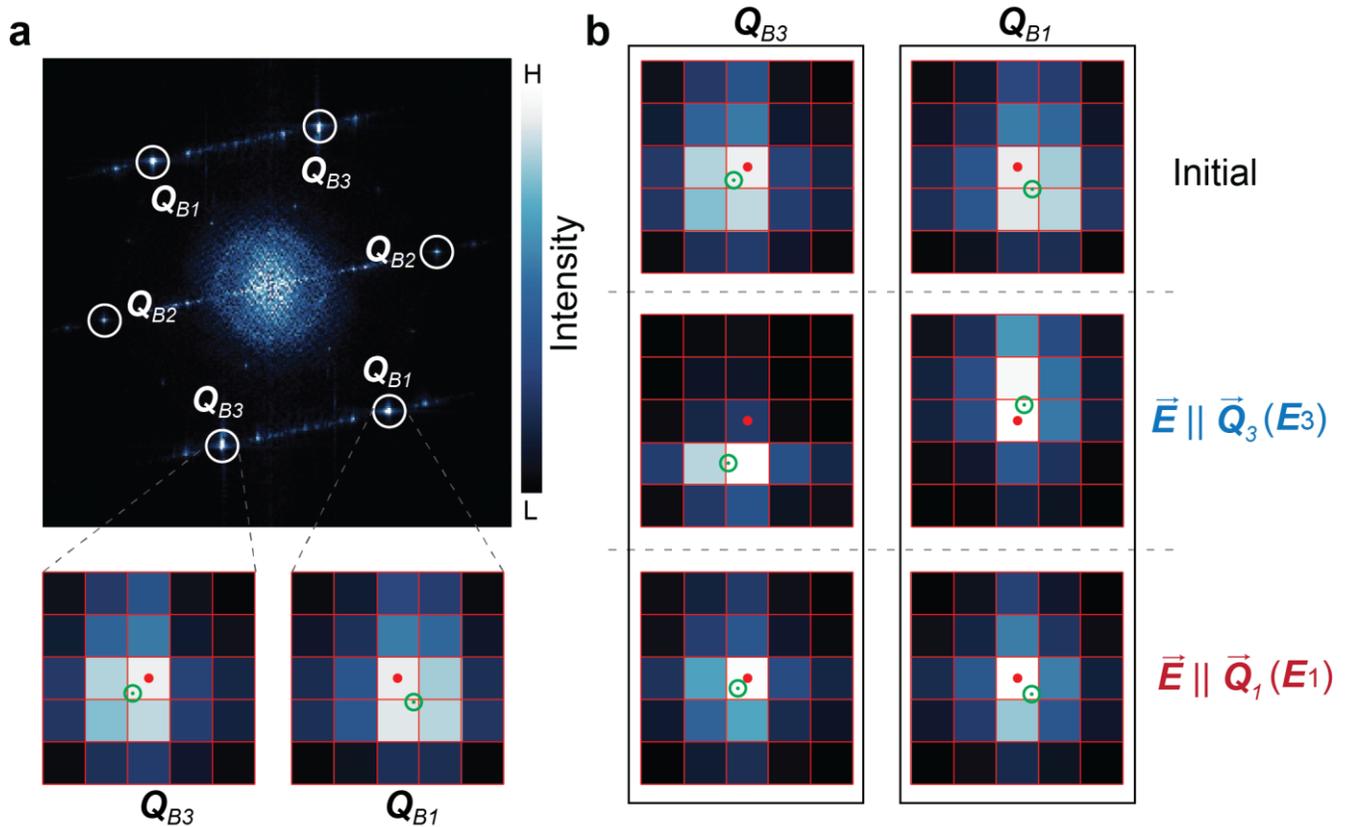

**Extended Data Fig. 8 | Identification of Bragg peak vector locations.**

**a**, Typical FT of the Sb layer before light illumination. Bragg peaks are clearly observed and highlighted with circles. Zoom-in images of these peaks along $Q_{B1}$ and $Q_{B3}$ directions reveal anisotropic intensity distribution in a 5-pixel x 5-pixel size in momentum space, thus enabling us to carry out the effective center of mass to identify the actual peak location (green circles). **b**, $Q_{B1}$ and $Q_{B3}$ peak locations during light illumination along either $E_1$ or $E_3$ directions, showing noticeable extending/shrinking peak location. The red dot in the middle of FT serves as reference for relative peak location change.

**Extended Data Fig. 9**

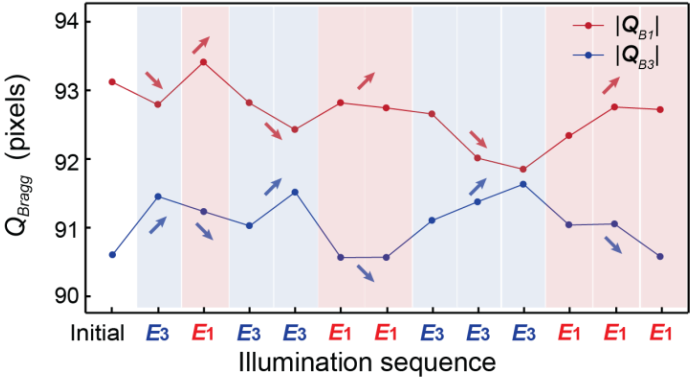

**Extended Data Fig. 9 | Absolute values of the Bragg vectors $|Q_{B1}|$ and $|Q_{B3}|$ with the laser polarization sequence.**

**Extended Data Fig. 10**

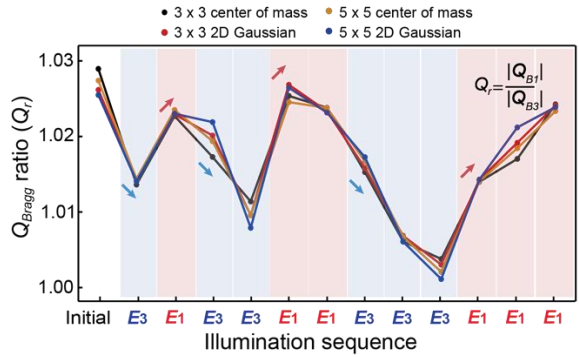

**Extended Data Fig. 10 | Bragg peak ratio ($Q_r$) along the laser polarization sequence with different grid sizes and peak position identification methods.**

Different grid sizes (3 x 3 and 5 x 5) and peak position identification methods (center of mass and 2D Gaussian) show similar results for the Bragg peak ratio $Q_r$. This demonstrates the robustness of the trends in the Bragg peak ratio for the laser/field sequence.

**Extended Data Fig. 11**

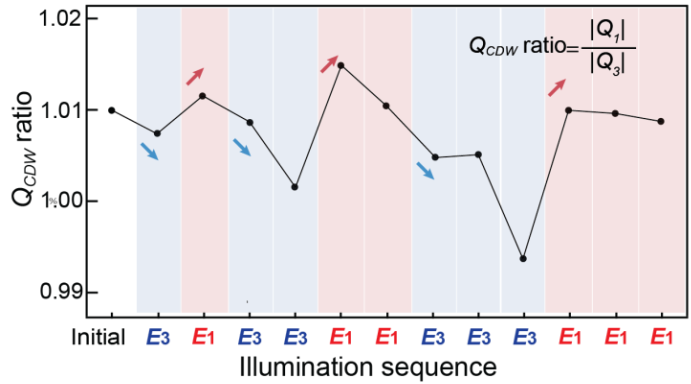

**Extended Data Fig. 11 | CDW peak vector ratios of $Q_1$ and $Q_3$ for an arbitrary laser illumination sequence.**

During the laser illumination sequence used in Fig. 3a and 3c, the CDW peak ratio shows the same pattern trend as the relative CDW intensity $I_r$ (Fig. 3a) and $Q_{Bragg}$ ratio $Q_r$ (Fig. 3c).

**Extended Data Fig. 12**

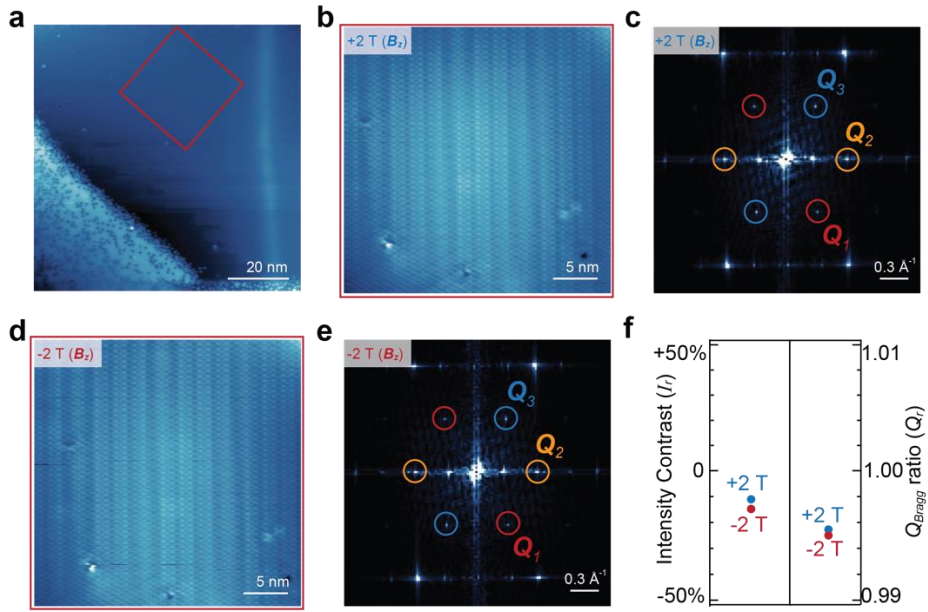

**Extended Data Fig. 12 | Data on a strained region where the CDW intensity contrast and Bragg vector ratio do not respond to the magnetic field.**

**a**, Large scale topography image (~ 87 nm x 87 nm) of the post-cleaned region. A wrinkle is clearly observed, indicating the presence of the strong residual in-plane strain in this region. **b,d**, Zoom-in topography image (30 nm x 30nm) of the same region under +2 T and -2 T magnetic field perpendicular to the sample surface. **c,e**, FTs of (b) and (d) show a robust intensity order between $Q_1$ and $Q_3$ with respect to the opposite directions of out-of-plane magnetic field. **f**, Intensity contrast plot ($Q_r$ in left panel) and Bragg vector ratio ($Q_r$ in right panel) from this region. The intensity contrast does not show a sign change between + 2 and -2 T field. The Bragg vector ratio shows a marginal change (≈ 0.0005) compared to the case (≈ 0.01, Fig. 4h) where there is a flip in the sign of the CDW intensity contrast (Fig. 4g).

**Extended Data Fig. 13**

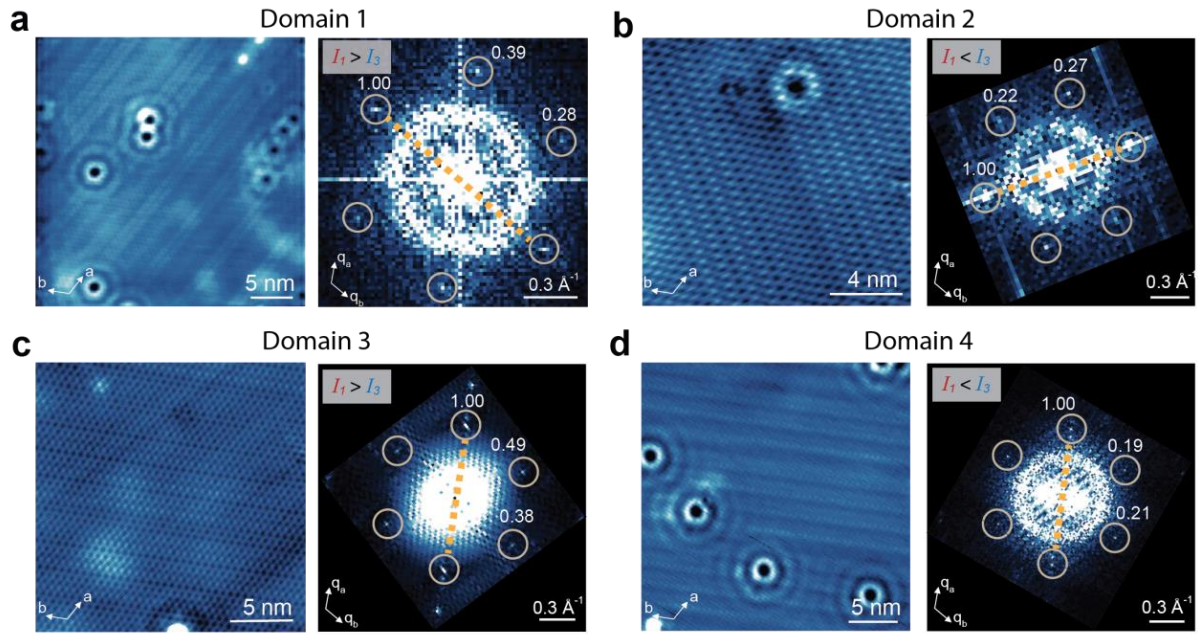

**Extended Data Fig. 13 | Survey of types of CDW domains from the same sample.**

For the proposed *Cmmm* magnetic space group of the CDW state, there exist three rotation symmetry-breaking directions and two time-reversal related partners for each of the three. Thus, six domain states are expected. Here in one single crystal sample, we present the topographic image and FFT of four different types of domains we have observed. The dotted yellow line represents the direction of the rotation symmetry breaking. The CDW intensity order is shown on the top left of each FFT.

**Extended Data Fig. 14**

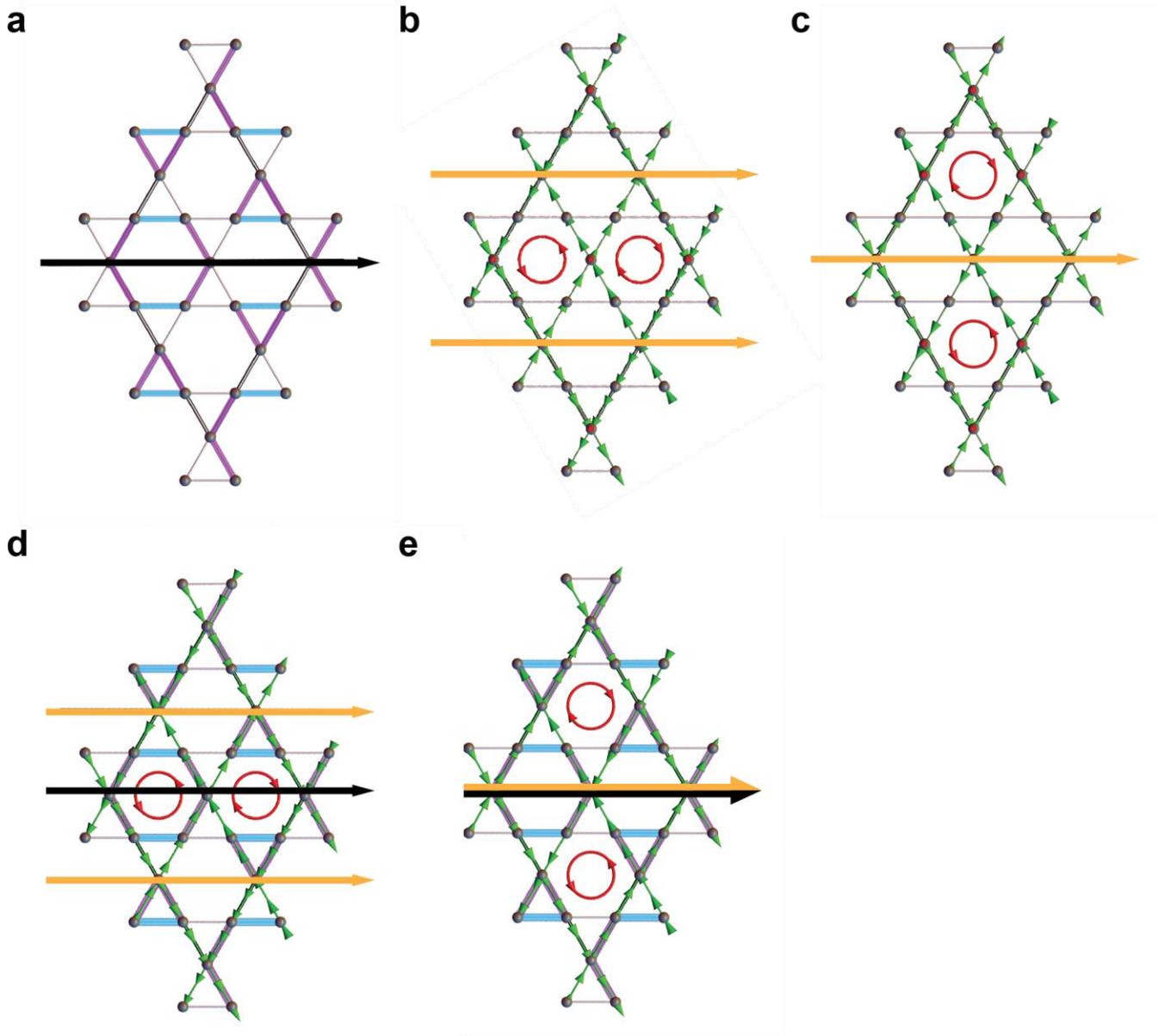

**Extended Data Fig. 14 | Vanadium bond order and loop current pattern and their rotation axis for different order parameter configurations.**

**a**, Vanadium bond-order pattern (colored bonds) for the 3Q "real" CDW order parameter configuration $\mathbf{L} = (L, L', L)$ on a single kagome layer. The black line denotes the in-plane 2-fold rotation axis. **b**, Loop-current pattern (green arrows) for the 2Q "imaginary" CDW order parameter configuration $\mathbf{\Phi} = (\Phi, 0, \Phi)$, whose relative phase between its non-zero components is trivial. The 2-fold rotation axis is denoted as the orange lines. The closed current loops are denoted as red circles. **c**, Loop-current pattern for the 2Q "imaginary" CDW order parameter configuration $\boldsymbol{\Phi} = (\Phi, 0, -\Phi)$. There exists a relative $\pi$ phase between the non-zero components, which changes the location of the 2-fold rotation axis. **d**, An overlay view of the $L_1 = L_3 \neq L_2$ and $\Phi_1 = \Phi_3$ orderings. In this case, bond-order and loop-current patterns have different rotation axis so that the 2-fold rotation is broken in the system, which allows a macroscopic out-of-plane magnetic dipole moment. **e**, An overlay view of the $L_1 = L_3 \neq L_2$ and $\Phi_1 = -\Phi_3$ orderings. In this case, both bond-order and loop-current patterns share the same rotation axis, so that the system possesses an in-plane 2-fold rotation axis, which forbids a macroscopic magnetic dipole moment. This configuration allows the system to have piezo-magnetic effect even in the absence of a macroscopic magnetic dipole moment.